\documentstyle[graphicx,times]{mn}

\newif\ifAMStwofonts
\ifoldfss
  \newcommand{\rmn}[1] {{\rm #1}}

  \ifCUPmtlplainloaded \else
    \NewTextAlphabet{textbfit} {cmbxti10} {}
    \NewTextAlphabet{textbfss} {cmssbx10} {}
    \NewMathAlphabet{mathbfit} {cmbxti10} {} % for math mode
    \NewMathAlphabet{mathbfss} {cmssbx10} {} %  "   "    "
  \fi
  \ifAMStwofonts
    \ifCUPmtlplainloaded \else
      \NewSymbolFont{upmath} {eurm10}
      \NewSymbolFont{AMSa} {msam10}
      \NewMathSymbol{\upi}     {0}{upmath}{19}
      \NewMathSymbol{\umu}     {0}{upmath}{16}
      \NewMathSymbol{\upartial}{0}{upmath}{40}
      \NewMathSymbol{\leqslant}{3}{AMSa}{36}
      \NewMathSymbol{\geqslant}{3}{AMSa}{3E}

       \let\le=\leqslant
       \let\ge=\geqslant
    \fi
  \fi
\fi % End of OFSS

\ifnfssone
  \newmathalphabet{\mathit}
  \addtoversion{normal}{\mathit}{cmr}{m}{it}
  \addtoversion{bold}{\mathit}{cmr}{bx}{it}
  \newcommand{\rmn}[1] {\mathrm{#1}}

  \newmathalphabet{\mathbfit} % math mode version of \textbfit{..}
  \addtoversion{normal}{\mathbfit}{cmr}{bx}{it}
  \addtoversion{bold}{\mathbfit}{cmr}{bx}{it}
  \newmathalphabet{\mathbfss} % math mode version of \textbfss{..}
  \addtoversion{normal}{\mathbfss}{cmss}{bx}{n}
  \addtoversion{bold}{\mathbfss}{cmss}{bx}{n}
  \ifAMStwofonts
    \ifCUPmtlplainloaded \else
      \UseAMStwoboldmath
      \makeatletter
      \new@mathgroup\upmath@group
      \define@mathgroup\mv@normal\upmath@group{eur}{m}{n}
      \define@mathgroup\mv@bold\upmath@group{eur}{b}{n}
      \edef\UPM{\hexnumber\upmath@group}
      \new@mathgroup\amsa@group
      \define@mathgroup\mv@normal\amsa@group{msa}{m}{n}
      \define@mathgroup\mv@bold\amsa@group{msa}{m}{n}
      \edef\AMSa{\hexnumber\amsa@group}
      \makeatother
      \mathchardef\upi="0\UPM19
      \mathchardef\umu="0\UPM16
      \mathchardef\upartial="0\UPM40
      \mathchardef\leqslant="3\AMSa36
      \mathchardef\geqslant="3\AMSa3E

       \let\le=\leqslant
       \let\ge=\geqslant
    \fi
  \fi
\fi % End of NFSS release 1

\ifnfsstwo
  \newcommand{\rmn}[1] {\mathrm{#1}}

  \DeclareMathAlphabet{\mathbfit}{OT1}{cmr}{bx}{it}
  \SetMathAlphabet\mathbfit{bold}{OT1}{cmr}{bx}{it}
  \DeclareMathAlphabet{\mathbfss}{OT1}{cmss}{bx}{n}
  \SetMathAlphabet\mathbfss{bold}{OT1}{cmss}{bx}{n}
  \ifAMStwofonts
    \ifCUPmtlplainloaded \else
      \DeclareSymbolFont{UPM}{U}{eur}{m}{n}
      \SetSymbolFont{UPM}{bold}{U}{eur}{b}{n}
      \DeclareSymbolFont{AMSa}{U}{msa}{m}{n}
      \DeclareMathSymbol{\upi}{0}{UPM}{"19}
      \DeclareMathSymbol{\umu}{0}{UPM}{"16}
      \DeclareMathSymbol{\upartial}{0}{UPM}{"40}
      \DeclareMathSymbol{\leqslant}{3}{AMSa}{"36}
      \DeclareMathSymbol{\geqslant}{3}{AMSa}{"3E}

       \let\le=\leqslant
       \let\ge=\geqslant
    \fi
  \fi
\fi % End of NFSS release 2

\ifCUPmtlplainloaded \else
  \ifAMStwofonts \else % If no AMS fonts
    \def\upi{\pi}
    \def\umu{\mu}
    \def\upartial{\partial}
  \fi
\fi

\title[Structure of Fornax \& Sgr globular clusters]
  {Surface brightness profiles and structural parameters for globular clusters in the Fornax and Sagittarius dwarf spheroidal galaxies}
\author[A.~D.~Mackey \& G.~F.~Gilmore]
  {A.~D.~Mackey$^1$\thanks{E-mail: dmackey@ast.cam.ac.uk}
  and G.~F.~Gilmore$^1$\\
  $^1$Institute of Astronomy, University of Cambridge, Madingley Road,
  Cambridge CB3 0HA}
\date{Accepted --. Received --}
\pagerange{000--000}
\pubyear{2002}

\def\LaTeX{L\kern-.36em\raise.3ex\hbox{a}\kern-.15em
    T\kern-.1667em\lower.7ex\hbox{E}\kern-.125emX}

\begin{document}

\label{firstpage}

\maketitle

\begin{abstract}
We present radial surface brightness profiles for all five globular 
clusters in the Fornax dwarf spheroidal galaxy, and for the four present
members of the Sagittarius dwarf spheroidal galaxy. These profiles
are derived from archival {\em Hubble Space Telescope} observations,
and have been calculated using the same techniques with which we measured
profiles in our previous studies of LMC and SMC clusters (Mackey \&
Gilmore 2002a,b), apart from some small modifications. From the surface 
brightness profiles, we have determined structural parameters for each 
cluster, including core radii and luminosity and mass estimates. We also 
provide a brief summary of literature measurements of other parameters for
these clusters, including their ages, metallicities and distances.

Our core radius measurements are mostly in good agreement with 
those from previous lower resolution studies, although for several 
clusters our new values are significantly different. The profile
for Fornax cluster 5 does not appear to be well fit by a King-type 
model and we suggest that it is a post core-collapse candidate. We 
examine the distribution of cluster core radii in each of the two 
dwarf galaxy systems, and compare these with the distribution of core 
radii for old LMC clusters. The three distributions match within the 
limits of measurement errors and the small sample sizes. We discuss the 
implications of this in the context of the radius-age trend we have 
previously highlighted for the Magellanic Cloud clusters.
\end{abstract}

\begin{keywords}
galaxies: star clusters -- globular clusters: general -- galaxies: individual: Fornax dwarf spheroidal, Sagittarius dwarf spheroidal -- Local Group -- stars: statistics
\end{keywords}

\section{Introduction}
Recently, we (Mackey \& Gilmore 2002a,b; hereafter Paper I and Paper II
respectively) presented high resolution surface brightness profiles for 
a large sample of clusters in the Large and Small Magellanic Clouds 
(LMC and SMC respectively), as measured with the Wide Field Planetary 
Camera 2 (WFPC2) on the {\em Hubble Space Telescope} ({\em HST}). 
These two galaxies are locally unique in containing
star clusters of comparable masses to the Milky Way's globular clusters
(see Papers I and II) but with ages ranging from the very newly 
formed ($\tau \sim 10^6$ yr) to those coeval with the oldest Galactic
globulars ($\tau \ge 10^{10}$ yr). The LMC and SMC systems therefore
permit direct observational studies of globular cluster evolution. 

One aspect of this is reflected in the trend in cluster structure we
highlighted in Papers I and II. When core radius ($r_c$) is plotted
as a function of age, clusters in both systems follow a distinct
distribution. The youngest clusters all have compact cores 
($r_c \sim 1-2$ pc), with the upper envelope of $r_c$ systematically 
increasing with age so that the oldest clusters exhibit a wide spread in 
their core sizes ($0 \le r_c \le 8$ pc). In Paper I, we showed that this
trend is almost certainly not the result of a selection effect or
a correlation between other intrinsic properties of the clusters. 
Furthermore, de Grijs et al. \shortcite{richard} showed that the 
radius-age distribution is not due to the comparison of profiles 
measured from different stellar masses -- that is, the reduced 
spread in $r_c$ for the youngest clusters cannot be a result of their 
profiles being dominated by a few high luminosity mass-segregated stars, 
because if only stars in the mass range $\sim 0.8 - 1.0 M_{\odot}$ are 
used to repeat the measurements (for a subsample of the clusters) the
radius-age distribution persists. It therefore seems that this observed
distribution represents some genuine form of time dependent structural 
alteration in clusters in both Magellanic Clouds.

The means by which some clusters obtain significantly expanded cores
while others do not remains unclear. In Paper I, we speculated on several
physical processes which are known to alter the structure of stellar
clusters, including the effects of mass loss due to stellar evolution
and the regulatory influence of the initial mass function (IMF),
the effects of heating due to three-body encounters and a significant
population of binary stars, the possibility of extreme tidal 
effects from the parent galaxies, and the mergers of pairs of 
gravitationally bound clusters. However, for the reasons discussed in
Paper I, none of these mechanisms seems to provide a viable means of 
reproducing the observed trend. Wilkinson et al. \shortcite{mark} have 
attacked the problem from a computational point of view. Their $N$-body
simulations have shown that it does indeed seem unlikely that significant
populations of binary stars or extreme tidal influences can
cause the necessary core expansion. This has led them to suggest that
perhaps the observed trend reflects changing formation conditions
rather than excessively dynamic cluster structures -- an idea which we 
are presently following up via additional simulations.

A second avenue of exploration is to observe if the radius-age
trend is present in any other globular cluster systems or whether it
is unique to the atypical Magellanic systems, thereby shedding light
on the possible evolutionary (or formation) conditions which might
lead to the $r_c$ distribution. There are only three
other globular cluster systems which are near enough to allow study
at sufficient resolution, and whose cluster censuses are (mostly)
complete. These are the Milky Way globular clusters, and those of the 
Fornax and Sagittarius dwarf spheroidal (dSph) galaxies. The Galactic 
globular cluster system is extensive, with $\sim 150$ members, and shows 
evidence for a complicated history of both formation and accretion of 
clusters (see e.g., Zinn \shortcite{zinn}). A detailed study of this 
system is therefore beyond the scope of this paper, but will be presented 
in the future (Mackey \& Gilmore, in prep.).

The Fornax and Sagittarius dSph galaxies are the two most massive
of the $\sim 11$ dwarf galaxies associated with the Milky Way, and
the only two to possess globular clusters (see e.g., Forbes et al. 
\shortcite{forbes}). Fornax contains five, while Sagittarius has
four definite members, plus at least one (Pal 12, see e.g., 
Mart\'{\i}nez-Delgado et al. \shortcite{pala}) probably previously 
associated. Unlike the Magellanic globular clusters however, the 
Fornax and Sagittarius clusters are exclusively old ($\sim 7 \le \tau \le 15$ 
Gyr). It is therefore not possible to directly observe any radius-age
trend for these two systems; however, we can measure the 
distribution of core radii and match it against that for the oldest
Magellanic clusters  -- thereby observing whether clusters with 
expanded cores are present in the two systems, and in what proportion.

To this end, we have located WFPC2 observations of the Fornax and
Sagittarius globular clusters in the {\em HST} archive, 
and put these data through the same reduction and
surface brightness profile construction pipelines we established for
the Magellanic Cloud clusters. This extends our group of directly 
comparable sets of measurements to all four of the closest globular
cluster systems external to the Milky Way. We describe the data 
and list a set of parameters we have compiled from the literature
for each cluster in Section 2, and briefly reiterate the reduction 
procedures in Section 3, as well as describing several alterations to 
this process which were required in order to deal with some of the 
characteristics of the new observations.
In Section 4 we present our results, and discuss these in the context
of the old Magellanic clusters in Section 5.
The results from the present study (Tables \ref{fornaxdata}, 
\ref{sgrdata}, \ref{ages}, \ref{distred}, \ref{params}, and 
\ref{luminmass}) together with the surface brightness profiles, are 
available on-line at 
{\em http://www.ast.cam.ac.uk/STELLARPOPS/dSph\_clusters/}.

\section{The cluster sample}
\label{sample}

\subsection{Observations}
\begin{table*}
\begin{minipage}{174mm}
\caption{Observation details for the Fornax dSph globular clusters.}
\begin{tabular}{@{}lccccccccccccc}
\hline \hline
Cluster & Program & & \multicolumn{5}{c}{Principal Frames$^{a,b}$} & & \multicolumn{5}{c}{Secondary Frames$^{a,b}$} \\
\cline{4-8}  \cline{10-14} \vspace{-3mm} \\
Name & ID & & Filter & Data-group & Date & $N_{e}$ & Time (s) & & Filter & Data-group & Date & $N_{e}$ & Time (s) \\
\hline
Fornax 1 & 5917 & & F555W & u30m010eb & 04/06/1996 & 14 & 5640 & & F814W & u30m010ib & 04/06/1996 & 16 & 7720 \\
Fornax 2 & 5917 & & F555W & u30m020eb & 06/06/1996 & 14 & 5640 & & F814W & u30m020ib & 06/06/1996 & 16 & 7720 \\
Fornax 3 & 5917 & & F555W & u30m030eb & 05/06/1996 & 14 & 5518 & & F814W & u30m030ib & 05/06/1996 & 16 & 7720 \\
Fornax 4 & 5637 & & F555W & u2lb0205b & 10/03/1995 & 3 & 2400 & & F814W & u2lb0203b & 10/03/1995 & 3 & 2400 \\
Fornax 5 & 5917 & & F555W & u30m040eb & 04/06/1996 & 14 & 5640 & & F814W & u30m040ib & 04/06/1996 & 16 & 7720 \\
\hline
\label{fornaxdata}
\end{tabular}
\medskip
\\
$^{a}$ $N_{e}$ is the total number of useful frames in the listed data-group. The F555W data-group for Fornax 3 has one exposure which is shorter than was intended ($378$s instead of $500$s; see below). \\
$^{b}$ The column labelled ``Time'' shows the total exposure time for a given data-group -- that is, the sum of the exposure times for each of the $N_{e}$ useful frames in that data-group. Not all frames had equal individual exposure times. The combinations are as follows: {\em (a)} Fornax 1, Fornax 2, \& Fornax 5: F555W $= 3\times 600$s, $4\times 500$s, $3\times 400$s, $4\times 160$s; F814W $= 2\times 900$s, $6\times 700$s, $2\times 500$s, $6\times 120$s; {\em (b)} Fornax 3: F555W $= 3\times 600$s, $3\times 500$s, $3\times 400$s, $1\times 378$s, $4\times 160$s; F814W $= 2\times 900$s, $6\times 700$s, $2\times 500$s, $6\times 120$s; {\em (c)} Fornax 4: F555W $= 2\times 1100$s, $1\times 200$s; F814W $= 2\times 1100$s, $1\times 200$s. \\
\end{minipage}
\end{table*}

\begin{table*}
\begin{minipage}{160mm}
\caption{Observation details for the Sagittarius dSph globular clusters.}
\begin{tabular}{@{}lccccccccccc}
\hline \hline
Cluster & Program & & \multicolumn{4}{c}{Principal Frame} & & \multicolumn{4}{c}{Secondary Frame} \\
\cline{4-7}  \cline{9-12} \vspace{-3mm} \\
Name & ID & & Filter & Data set & Date & Time (s) & & Filter & Data set & Date & Time (s) \\
\hline
M54 & 6701 & & F555W & u37ga409r & 30/08/1999 & 300 & & F814W & u37ga401r & 30/08/1999 & 300 \\
Terzan 7 & 6701 & & F555W & u37g020br & 18/03/1997 & 300 & & F814W & u37g0202r & 18/03/1997 & 300 \\
Terzan 8 & 6701 & & F555W & u37g1307r & 26/03/1999 & 300 & & F814W & u37g1306r & 26/03/1999 & 260 \\
Arp 2 & 6701 & & F555W & u37g0107m & 11/05/1997 & 300 & & F814W & u37g0101m & 11/05/1997 & 300 \\
\hline
\label{sgrdata}
\end{tabular}
\end{minipage}
\end{table*}

In order to ensure results from this reduction and analysis were
directly comparable to the results from Papers I and II, we required
observations in the same format as in these two studies -- that is,
in the general form of a cluster observed once (or in selected cases,
multiple times) through the F555W filter, and once (or again, multiple
times) through a second filter. Two {\em HST} programs have observed 
Fornax dSph globular clusters with WFPC2 -- program
5917 (Fornax 1, 2, 3, and 5) and program 5637 (Fornax 4). As described
below in Section \ref{reduction}, because of the large distance of
the Fornax system we required multiple exposures of each cluster,
and hence we downloaded all the available data from these programs.
The full data set is listed in Table \ref{fornaxdata}. All clusters
were observed through the F555W and F814W filters. For the clusters
observed in program 5917, typical total exposure times were 5640 seconds
in F555W and 7720 seconds in F814W, with 14 and 16 frames of differing
individual exposure times in the two data-groups respectively. 
The cluster Fornax 4 was observed in program 5637 and has somewhat
more limited data -- total exposure times of 2400 seconds in both filters,
and only 3 frames per data-group. Again, the individual frames have
varying exposure durations. These are fully described
in the notes to Table \ref{fornaxdata}.

The four Sagittarius dSph globular clusters have only been 
targeted by one {\em HST} WFPC2 program -- number 6701. Again, the 
observations were made using the F555W and F814W filters. For these four
clusters (again as described in Section \ref{reduction}) we only 
required one frame in each colour -- we selected those of exposure
duration 300 seconds (except for Terzan 8 through the F814W filter, 
for which only a 260 second exposure was usable). There were no 
significantly shorter or longer exposures available in the archive. 

Finally, it is worth noting that it has recently been plausibly 
demonstrated that the young outer halo globular cluster Palomar 12 is a 
former member of the Sagittarius dwarf, taking its census of clusters to 
five \cite{pala,palb}. Unfortunately, Pal 12 has has not been observed 
by {\em HST} and therefore could not be added to the present study.

\subsection{Literature data}
In the interests of presenting a detailed study of the two dSph cluster
systems, we have compiled literature data for the positions, ages,
metallicities, distances, and reddenings of all nine clusters. The
position, age, and metallicity data are presented in Table \ref{ages},
and the distance and reddening data in Table \ref{distred}. Like the
literature compilations in Papers I and II, this tabulation is not
intended to be an exhaustive study of the entire available literature --
its purpose is rather to provide a consistent set of measurements
for use in this and future work. 

\subsubsection{Cluster names and positions}
We have taken the most common names for the nine clusters from the
Simbad Astronomical Database ({\em http://simbad.u-strasbg.fr/}). The
Fornax clusters are labelled as Fornax $1 - 5$, according to the 
notation of Hodge \shortcite{hodge}. Cluster Fornax 3 also has an
NGC designation -- it is NGC 1049. M54 is the brightest of the 
Sagittarius clusters, and is also known as NGC 6715. The remaining three
clusters are of significantly lower surface density and do not have 
NGC entries. The names of clusters Terzan 7 and Terzan 8 refer to the 
catalogue of Terzan \shortcite{terzan}, while that for cluster Arp 2 
refers to the list by Arp \shortcite{arp}. 

All positions are also taken from Simbad.
For the Sagittarius clusters, these positions correspond to those
presented by Harris \shortcite{harris} (1999 update). Using these
positions we calculated the angular separations $R_{ang}$ of the 
clusters from the centres of their respective parent galaxies. For the
Fornax dSph we adopted the optical centre from Simbad, at
$\alpha = 02^{h}39^{m}59\fs3$, $\delta = -34\degr 26\arcmin 57\arcsec$ 
(J2000.0) -- this is consistent with the centroids derived by Demers 
et al. \shortcite{demersb}, and Irwin \& Hatzidimitriou 
\shortcite{dwarfpars}. For the Sagittarius dSph, it is difficult to 
define an optical centre because of the severely disrupted nature of the 
galaxy. We have adopted the centre to be the location of the densest 
clump of stars, upon which M54 is exactly superposed 
\cite{sgrnature,ibatab}, at 
$\alpha = 18^{h}55^{m}03\fs28$, $\delta = -30\degr 28\arcmin 42\farcs6$
(J2000.0) -- indeed, it has been suggested (e.g., Layden \& Sarajedini
\shortcite{laysar}) that Sagittarius is a nucleated dwarf with M54 as
its nucleus. Using $R_{ang}$ and the distances from Table \ref{distred},
we have also calculated $R_{lin}$ -- the linear separation of a given
cluster from the centre of its parent galaxy. Because we have individual
distance moduli for the Sagittarius clusters (Table \ref{distred}), 
these calculations naturally account for the line-of-sight depth of 
this system.

\subsubsection{Cluster ages}
\label{clusterages}
As far as we are aware, the only available studies concerning the 
relative ages of the Fornax globular clusters are those by 
Buonanno and collaborators (Buonanno et al. 1998b,1999), who use the same 
{\em HST} images from Table
\ref{fornaxdata} to construct colour magnitude diagrams (CMDs) from
which they derive relative ages and metallicities for the clusters. 
Clusters 1, 2, 3, and 5 appear coeval with the oldest Galactic globulars,
and we adopt absolute ages of $14.6 \pm 1.0$ Gyr (for consistency with
the oldest Sagittarius clusters -- see below). Cluster 4 is $\sim 3$
Gyr younger, and we adopt an absolute age of $11.6 \pm 1.0$ Gyr.

The Sagittarius clusters are more extensively studied. Layden \& 
Sarajedini \shortcite{laysar} have calculated relative and absolute ages 
for each, and we adopt their ages for M54 and Terzan 8 in particular, as 
$14.7 \pm 0.5$ Gyr and $14.5 \pm 0.8$ Gyr respectively. They show these 
clusters to be coeval with typical Galactic globulars of comparable
metallicity (e.g., M92 and M5). We therefore used these two ages to
estimate absolute ages for Fornax clusters 1, 2, 3, and 5, as listed
above. We note that our adopted age for Terzan 8 is consistent with the
work of Montegriffo et al. \shortcite{monte}, who also find this cluster 
to be coeval with typical metal-poor Galactic globular clusters. 

Terzan 7 and Arp 2 appear younger than Terzan 8 and M54, and there
are several bodies of work concerning these two clusters. For Arp 2
we directly average the results of Buonanno \shortcite{buonages}
(who derive an age of $-1.6 \pm 1.6$ Gyr relative to the oldest 
globular clusters -- i.e., in this case M54 and Terzan 8), Layden
\& Sarajedini \shortcite{laysar} ($13.1 \pm 0.9$ Gyr), and Salaris
\& Weiss \shortcite{salaris} ($11.5 \pm 1.4$ Gyr), which are all
in good agreement, to obtain $12.5 \pm 1.0$ Gyr. Montegriffo et al.
\shortcite{monte} find an age of $-4.4 \pm 2.0$ Gyr relative to Terzan 8,
and we have not included this significantly younger age in the average.
For Terzan 7 however, all four studies obtain consistent age estimates
-- respectively, $-7.6 \pm 1.7$ Gyr relative to the oldest Galactic 
clusters; $8.3 \pm 1.8$ Gyr; $7.5 \pm 1.4$ Gyr; and $-6.9 \pm 2.0$ Gyr
relative to Terzan 8. We directly average these results to obtain
an age estimate of $7.6 \pm 1.0$ Gyr for Terzan 7 -- one of the 
youngest globular clusters.

\begin{table*}
\begin{minipage}{175mm}
\caption{Literature nomenclature, position, age and metallicity data for the cluster sample.}
\begin{tabular}{@{}lccccccccc}
\hline \hline
Principal & \multicolumn{2}{c}{Position (J2000.0)} & $R_{ang}$ & $R_{lin}$ & $\tau$ & $\log \tau$ & Age$^{d}$ & Metallicity & Metallicity$^{d}$ \vspace{0.5mm} \\
Name$^{a}$ & $\alpha$ & $\delta$ & $(\degr)^{b}$ & (kpc)$^{c}$ & (Gyr) & (yr) & References & $[$Fe$/$H$]$ & References \\
\hline
Fornax 1 & $02^{h}37^{m}02\fs1$ & $-34\degr 11\arcmin 00\arcsec$ & $0.67$ & $1.60$ & $14.6 \pm 1.0$ & $10.16 \pm 0.03$ & $6$ & $-2.20 \pm 0.20$ & $6\,\,(2,9,12)$ \\
Fornax 2 & $02^{h}38^{m}40\fs1$ & $-34\degr 48\arcmin 05\arcsec$ & $0.44$ & $1.05$ & $14.6 \pm 1.0$ & $10.16 \pm 0.03$ & $6$ & $-1.78 \pm 0.20$ & $6\,\,(1,2)$ \\
Fornax 3 & $02^{h}39^{m}52\fs5$ & $-34\degr 16\arcmin 08\arcsec$ & $0.18$ & $0.43$ & $14.6 \pm 1.0$ & $10.16 \pm 0.03$ & $6$ & $-1.96 \pm 0.20$ & $6\,\,(2,10,12)$ \\
Fornax 4 & $02^{h}40^{m}09^{s}$ & $-34\degr 32\arcmin 24\arcsec$ & $0.10$ & $0.24$ & $11.6 \pm 1.0$ & $10.06 \pm 0.04$ & $7$ & $-1.9 \pm 0.2$ & $7\,\,(1,10)$ \\
Fornax 5 & $02^{h}42^{m}21\fs15$ & $-34\degr 06\arcmin 04\farcs7$ & $0.60$ & $1.43$ & $14.6 \pm 1.0$ & $10.16 \pm 0.03$ & $6$ & $-2.20 \pm 0.20$ & $6\,\,(2,10)$ \\
\hline
M54 & $18^{h}55^{m}03\fs28$ & $-30\degr 28\arcmin 42\farcs6$ & $0.00$ & $0.00$ & $14.7 \pm 0.5$ & $10.17 \pm 0.02$ & $13$ & $-1.79 \pm 0.08$ & $17\,\,(8,11,19)$ \\
Terzan 7 & $19^{h}17^{m}43\fs7$ & $-34\degr 39\arcmin 27\arcsec$ & $6.26$ & $4.85$ & $7.6 \pm 1.0$ & $9.88^{+0.05}_{-0.06}$ & $5,13,14,16$ & $-0.82 \pm 0.15$ & $18\,\,(3,8,11)$ \\
Terzan 8 & $19^{h}41^{m}45\fs0$ & $-34\degr 00\arcmin 01\arcsec$ & $10.30$ & $4.92$ & $14.5 \pm 0.8$ & $10.16^{+0.02}_{-0.03}$ & $13\,\,(14)$ & $-1.99 \pm 0.08$ & $8\,\,(14,15)$ \\
Arp 2 & $19^{h}28^{m}44\fs1$ & $-30\degr 21\arcmin 14\arcsec$ & $7.27$ & $3.80$ & $12.5 \pm 1.0$ & $10.10^{+0.03}_{-0.04}$ & $5,13,16\,\,(14)$ & $-1.84 \pm 0.09$ & $18\,\,(4,8,11)$ \\
\hline
\label{ages}
\end{tabular}
\medskip
\\
Reference list (see also text): 1. Beauchamp et al. \shortcite{beau}; 2. Buonanno et al. \shortcite{buonforold}; 3. Buonanno et al. \shortcite{buonmeta}; 4. Buonanno et al. \shortcite{buonmetb}; 5. Buonanno et al. \shortcite{buonages}; 6. Buonanno et al. \shortcite{buonfora}; 7. Buonanno et al. \shortcite{buonforb}; 8. Da Costa \& Armandroff \shortcite{garysgr}; 9. Demers et al. \shortcite{demersa}; 10. Dubath et al. \shortcite{dubath}; 11. Harris \shortcite{harris} (1999 update); 12. J\o rgensen \& Jimenez \shortcite{jj}; 13. Layden \& Sarajedini \shortcite{laysar}; 14. Montegriffo et al. \shortcite{monte}; 15. Ortolani \& Gratton \shortcite{orto}; 16. Salaris \& Weiss \shortcite{salaris}; 17. Sarajedini \& Layden \shortcite{sarlaya}; 18. Sarajedini \& Layden \shortcite{sarlayb}; 19. Zinn \& West \shortcite{zinnwest}. \\
$^{a}$ Fornax 3 is also known as NGC 1049, while M54 is also known as NGC 6715. \\
$^{b}$ Angular separation relative to the optical centre of {\em (a)} the Fornax dSph, at $\alpha = 02^{h}39^{m}59\fs3$, $\delta = -34\degr 26\arcmin 57\arcsec$ (J2000.0)\ (Simbad Astronomical Database, see also Demers et al. \shortcite{demersb}; Irwin \& Hatzidimitriou \shortcite{dwarfpars}); {\em (b)} the Sagittarius dSph, taken to be the position of M54 (see e.g., Layden \& Sarajedini \shortcite{laysar}; Ibata et al. \shortcite{ibatab}; Ibata et al. \shortcite{sgrnature}). \\
$^{c}$ Estimated linear separations to the optical centres of either the Fornax or Sagittarius dwarfs, calculated from $R_{ang}$ and the cluster distance moduli from Table \ref{distred}. We assume the distance to the centre of the Fornax dSph to be $\sim 137$ kpc \cite{buonforb}, and that for the Sagittarius dSph to be equivalent to the M54 distance from Harris \shortcite{harris}. \\
$^{d}$ Principal references (i.e., for the adopted metallicity or age value) are given first. In some cases these are followed by references in parenthesis, which show literature complementary to the principal reference, as described in the text. Entries with multiple non-bracketed references indicate that an average of the individual values from these respective sources has been adopted, again as described in the text. \\
\end{minipage}
\end{table*}

\subsubsection{Cluster metallicities}
There are several estimates of metallicity per cluster available in
the literature. For the Fornax clusters, mostly these are photometric
estimates from CMDs. For consistency, we have adopted the results of
the highest resolution CMD studies - those of Buonanno et al. 
\shortcite{buonfora,buonforb}. These are in good agreement with other
previous photometric measurements for Fornax 1, 2, 3, and 5 (e.g., 
Buonanno et al. \shortcite{buonforold}; Demers et al. \shortcite{demersa};
Beauchamp et al. \shortcite{beau}; J\o rgensen \& Jimenez \shortcite{jj}).
Dubath, Meylan \& Mayor \shortcite{dubath} provide spectroscopic
metallicity determinations for Fornax 3, 4, and 5. Again, those for
Fornax 3 and 5 are in good agreement with the photometric estimates. 
Cluster 4 however, shows a discrepancy between the photometric
metallicity determinations \cite{beau,buonforb} and those derived from 
spectroscopy \cite{dubath,beau} -- the measurements are 
[Fe/H] $\sim -1.9$ and [Fe/H] $\sim -1.3$ respectively. 
Cluster 4 is in a relatively dense region of 
the Fornax dwarf, and it is possible that field stars (which are measured
to have [Fe/H] $\sim -1.3$) have contaminated the spectroscopic 
measurements. However, Buonanno et al \shortcite{buonforb} note that a 
similar discrepancy is well known for several Galactic clusters -- in 
particular Rup 106 and Pal 12 -- and that it may be linked to the fact 
that these clusters appear to be [$\alpha$/Fe]-deficient relative to 
other globular clusters (see also Sarajedini \& Layden 
\shortcite{sarlayb} for a detailed discussion). It might be that Fornax 
4 is a similar case. 

Interestingly, the (young) Sagittarius clusters Terzan 7 and Arp 2
show a similar discrepancy between their photometrically and 
spectroscopically determined metallicities. For Terzan 7, the CMD study
of Sarajedini \& Layden \shortcite{sarlayb} measured
[Fe/H] $=-0.82 \pm 0.15$, which is consistent with the study of
Buonanno et al. \shortcite{buonmeta}. However, Da Costa \& Armandroff
\shortcite{garysgr} obtained [Fe/H] $=-0.36 \pm 0.09$ from spectroscopy
of the Ca {\sc ii} triplet. Similarly, for Arp 2 Sarajedini \& Layden 
\shortcite{sarlayb} measured [Fe/H] $=-1.84 \pm 0.09$ (in good agreement
with Buonanno et al. \shortcite{buonmetb}) while Da Costa \&
Armandroff \shortcite{garysgr} observed [Fe/H] $=-1.70 \pm 0.11$. These
discrepancies are in the same sense as those for Fornax 4, Rup 106, and
Pal 12 -- namely that the spectroscopic measurements suggest higher
metallicities than are consistent with the clusters' CMDs. For the 
present, we adopt, for internal consistency, the photometrically determined 
values; further observations will no doubt clarify the situation in the near
future. We feel that it is worth noting that all of the clusters
which show this discrepancy are conclusively linked with local dwarf
galaxies, except Rup 106. It would be worthwhile searching in the vicinity
of this cluster for any evidence of debris from a former parent galaxy, 
such as that found surrounding Pal 12 by Mart\'{\i}nez-Delgado et al. 
\shortcite{pala}.

Metallicity determinations for Terzan 8 are much simpler to interpret.
There is good agreement between the spectroscopic measurements by
Da Costa \& Armandroff \shortcite{garysgr} and the photometric estimates
of Montegriffo et al. \shortcite{monte} and Ortolani \& Gratton
\shortcite{orto}. For M54, again there appears a discrepancy between
the photometric measurement of Sarajedini \& Layden \shortcite{sarlaya}
and the spectroscopy of Da Costa \& Armandroff \shortcite{garysgr}.
However, like for the massive globular cluster $\omega$ Cen, Sarajedini
\& Layden \shortcite{sarlaya} have suggested that M54 may possess an
internal metallicity dispersion, of $\sigma($[Fe/H]$) = 0.16$ dex. 
Da Costa \& Armandroff conclude that if this is indeed the case then
the two measurements are consistent. As with the other clusters, we adopt
the photometric determination.

\subsubsection{Distances and reddenings}
\label{secred}
\begin{table*}
\begin{minipage}{151mm}
\caption{Literature distance and reddening data for the cluster sample.}
\begin{tabular}{@{}lccccccc}
\hline \hline
Cluster & Distance & Reference$^{b}$ & Distance & Scale Factor & $E(B-V)^{c}$ & $E(V-I)^{c}$ & Reference$^{b}$ \vspace{0.5mm} \\
 & Modulus$^{a}$ & & (kpc) & (arcsec pc$^{-1}$) & & & \\
\hline
Fornax 1 & $20.68 \pm 0.20$ & $7$ & $137\,\,(\pm 13)$ & $1.508\,\,(\pm 0.144)$ & $(0.04 \pm 0.05)$ & $0.05 \pm 0.06$ & $6$ \\
Fornax 2 & $20.68 \pm 0.20$ & $7$ & $137\,\,(\pm 13)$ & $1.508\,\,(\pm 0.144)$ & $(0.07 \pm 0.05)$ & $0.09 \pm 0.06$ & $6$ \\
Fornax 3 & $20.68 \pm 0.20$ & $7$ & $137\,\,(\pm 13)$ & $1.508\,\,(\pm 0.144)$ & $(0.04 \pm 0.05)$ & $0.05 \pm 0.06$ & $6$ \\
Fornax 4 & $20.68 \pm 0.20$ & $7$ & $137\,\,(\pm 13)$ & $1.508\,\,(\pm 0.144)$ & $(0.12 \pm 0.05)$ & $0.15 \pm 0.06$ & $7$ \\
Fornax 5 & $20.68 \pm 0.20$ & $7$ & $137\,\,(\pm 13)$ & $1.508\,\,(\pm 0.144)$ & $(0.06 \pm 0.05)$ & $0.08 \pm 0.06$ & $6$ \\
\hline
M54 & $17.17 \pm 0.15$ & $11$ & $27.2\,\,(\pm 1.9)$ & $7.583\,\,(\pm 0.532)$ & $(0.14 \pm 0.02)$ & $0.18 \pm 0.02$ & $13$ \\
Terzan 7 & $16.83 \pm 0.15$ & $11$ & $23.2\,\,(\pm 1.6)$ & $8.891\,\,(\pm 0.616)$ & $0.07 \pm 0.03$ & $(0.09 \pm 0.04)$ & $18$ \\
Terzan 8 & $17.08 \pm 0.15$ & $11$ & $26.0\,\,(\pm 1.8)$ & $7.933\,\,(\pm 0.552)$ & $0.12 \pm 0.03$ & $(0.15 \pm 0.04)$ & $14$ \\
Arp 2 & $17.28 \pm 0.15$ & $11$ & $28.6\,\,(\pm 2.0)$ & $7.212\,\,(\pm 0.507)$ & $0.10 \pm 0.02$ & $(0.13 \pm 0.03)$ & $18$ \\
\hline
\end{tabular}
\medskip
\\
$^{a}$ Dereddened visual distance modulus. For the Sagittarius clusters, we have taken the apparent distance moduli of Harris \shortcite{harris} and corrected these using the reddenings of columns 6 \& 7 and the extinction laws described in Section \ref{secred}. \\
$^{b}$ Numbers refer to entries in the Reference list at the end of Table \ref{ages}. \\
$^{c}$ Values in parenthesis have been calculated from the neighbouring column using the extinction laws described in Section \ref{secred}. \\
\label{distred}
\end{minipage}
\end{table*}

Accurate distances are necessary for converting measured parameters
from observational to physical units (e.g., arcseconds to parsecs),
as well as for estimating total cluster luminosities -- a task for
which we also require accurate reddenings. Since both calculations
were desirable for this study (see Section \ref{results}), we also
compiled distance and reddening data from the literature, as listed
in Table \ref{distred}. For the Fornax clusters, we adopted a single
(dereddened) distance modulus of $20.68 \pm 0.20$, as suggested from the 
extensive discussion of Buonanno et al. \shortcite{buonforb}. This 
corresponds to a distance of $137 \pm 13$ kpc, consistent with other
literature measurements (e.g., Buonanno et al. \shortcite{buonforold};
Demers et al. \shortcite{demersa}). The high resolution CMD studies 
of Buonanno and collaborators \cite{buonfora,buonforb} also provide
$E(V-I)$ colour excesses, as listed in Table \ref{distred}. 

For completeness, we would like to include estimates of the $E(B-V)$
colour excesses for these clusters. This requires some knowledge of
interstellar extinction laws, which are also required for calculating
dereddened distance moduli for the Sagittarius clusters (see below).
We take the definition of the colour excess as 
$E(\lambda_1 - \lambda_2) = A_{\lambda_1} - A_{\lambda_2}$, where
$\lambda_1$ and $\lambda_2$ are the passbands being considered, and
the $A_\lambda$ values are the extinctions in these passbands. Following
Cardelli, Clayton \& Mathis \shortcite{red}, we can also define
$c = A_{\lambda_2} / A_{\lambda_1}$, which means that in our present
case (where $V$ and $I$ are our two passbands) we have
$A_V = (1-c)^{-1} E(V-I)$. Interpolating in Table 3 of Cardelli, Clayton
\& Mathis \shortcite{red}, we find that $c\sim 0.578$, assuming that the 
centre of the $I$ passband is 814 nm. Therefore, we obtain 
$A_V = 2.37E(V-I)$. Cardelli et al. make the assumption that 
$A_V = 3.1E(B-V)$, which implies that $E(V-I)=1.31E(B-V)$. These relations
allow us to compute reddenings and colour excesses as appropriate -- this
is also important in the luminosity calculations of Section 
\ref{lumandmass}.

For the Sagittarius clusters, we adopt the individual apparent distance 
moduli from Harris \shortcite{harris} (1999 update), and have dereddened 
them using the quoted colour excesses (also listed by Harris) and the
extinction laws described above to obtain the values in
Table \ref{distred}. The original references for the colour excesses
(all determined from high data-quality CMD studies) are as follows: 
M54, Layden \& Sarajedini \shortcite{laysar}; Terzan 8, Montegriffo et 
al. \shortcite{monte}; Terzan 7 and Arp 2, Sarajedini \& Layden
\shortcite{sarlayb}.

\section{Photometry and surface brightness profiles}
\label{reduction}
The data reduction, photometry, and construction of surface brightness
profiles followed almost exactly the procedures outlined in Papers I 
and II for the Magellanic Cloud clusters, and for a detailed description 
we refer the reader to these. However, the Fornax and Sagittarius 
globular clusters, at respectively twice and half the distances of the 
Magellanic clusters, presented their own set of challenges, and in
several places we altered the reduction procedure. For continuity, 
we therefore provide a brief summary of the overall reduction process 
below, and describe the details of the changes made to the procedure
of Papers I and II where applicable.

The first alteration came at the very start of the reduction process,
during the image preparation. As described in Papers I and II, for
the old ($\tau \sim 10^{10}$ yr) Magellanic Cloud clusters, it had
generally been sufficient to use one frame of several hundred seconds' 
exposure time, in each of two colours per cluster, to obtain the 
photometry for use in constructing the surface brightness profiles.
For the Sagittarius clusters, this was again sufficient. The Fornax 
clusters however, are more than twice as distant as the Magellanic Cloud 
clusters, and therefore appear considerably fainter and more crowded. 
For these clusters, it was necessary to take many frames per colour per 
cluster (Table \ref{fornaxdata}), and stack them together to obtain 
suitable photometry.

The data were first retrieved from the {\em HST} archive. As part of
the retrieval process, all exposures are reduced according to the 
standard WFPC2 pipeline, using the latest available calibrations. This
ensures that all the data has been treated using the calibration from a 
single epoch, and with the longest baseline. To obtain photometry
from these reduced frames, we used the HSTphot software of Dolphin
\shortcite{hstphot}, which is specifically tailored to measuring
WFPC2 images. First, we pre-treated each image with the HSTphot
preparation software, as described in Papers I and II. The Sagittarius 
data consisted of two frames (one F555W and one F814W) per cluster 
(see Table \ref{sgrdata}). On each frame, bad regions and pixels were 
masked using the STScI data quality images, cosmic rays were removed 
using a HSTphot routine based on the {\sc iraf} task {\sc crrej}, 
potential hot-pixels were removed using a $\sigma$-clipping algorithm, 
and a background image was prepared.

The Fornax data were more complicated, consisting of two data-groups
(one F555W and one F814W) of multiple frames per cluster. 
On each frame, bad regions and pixels 
were masked using the STScI data quality images, as usual. However,
next the procedure deviated from that described in Papers I and II. 
First, we aligned all the frames in a specific data-group using the 
{\sc iraf} task {\sc imalign}. This routine takes a base image and treats
each of the other frames as being offset from this base by a simple linear
$x$- and $y$-shift. This was perfectly suitable for the present data --
all frames in a specific data-group were imaged one after the other,
and any offsets were due to a deliberate dithering pattern of
several pixels in $x$ and/or $y$. There were no significant rotations
or higher order distortions between observations which needed to be
accounted for. Once registered, all the frames from a data-group 
were added together using the HSTphot routine {\em crclean}. This
routine cleans cosmic rays (again using an algorithm similar to
{\sc crrej}), then adds together the counts for a given pixel in all
images and scales for the total exposure time. This means that images
of multiple exposure times can be combined, with the resultant ``master''
frame having a much higher saturation level than any of the individual
images, and an effective exposure time which is the sum of that for
the individual images. For each colour for each cluster, we therefore
added long duration exposures to short duration exposures and
the resultant image had a very low faint limit but did not suffer badly
from crowding (except in the most central regions of Fornax 3, 4, and 5).
This process thereby mostly alleviated the two major problems caused 
by the large distance of the Fornax clusters.

We then cleaned each master frame with the hot-pixel algorithm and
determined a background image, just as in the usual procedure.
From this point on, the two master frames for a Fornax cluster were 
treated exactly as the two individual frames for the clusters described 
in Papers I and II and the Sagittarius clusters.

Photometric measurements were made using the HSTphot module 
{\em multiphot} in PSF fitting mode, with a minimum detection threshold
of $3\sigma$ above the local background. This routine aligns and 
solves the two (different colour) frames for a cluster simultaneously,
which eliminates the need for correlating and matching between object
lists, and is useful for keeping the photometry clean of spurious
(e.g., cosmic ray) detections -- only objects found in both frames were
kept. Parameters derived from the PSF fitting, such as the object 
classification, sharpness, and goodness-of-fit were used as described in 
Papers I and II to further clean the detection list of spurious and 
non-stellar objects. Photometry in both colours for the selected objects 
was corrected for geometric distortion, any filter-dependent plate scale 
changes, and the WFPC2 34th row defects and charge transfer inefficiency 
(using the calibration of Dolphin \shortcite{cte}). Any PSF residuals 
were accounted for and the final photometry corrected to a $0\farcs5$ 
aperture and the zero-points of Dolphin \shortcite{cte}.

With the photometry complete, for each cluster we calculated positions 
for all stars in corrected pixel coordinates (pixel coordinates 
corrected for geometric distortion and relative to the WFC2 origin) 
using the {\sc iraf stsdas} task {\sc metric}. We then determined the 
position of the cluster's central surface brightness peak, using the
random sampling algorithm described in Paper I. This centering is
performed using the photometry in both colours (a good consistency
check) and is typically repeatable to $\pm 10$ WFC pixels, or 
approximately $\pm 1$ second of arc.

In each colour, we measured the cluster's surface brightness profile via 
radial binning 
about the calculated centre. For each cluster, four different
sets of annuli were generated. In the case of the Fornax clusters and
M54, two sets were of narrow ($1.5\arcsec$ and $2\arcsec$) width and
extended to $\sim 20\arcsec$ and $\sim 30\arcsec$ respectively, while
the remaining two sets were wider ($3\arcsec$ and $4\arcsec$) and
extended as far as possible (typically $\sim 80\arcsec$). This is just
as described for the Magellanic Cloud clusters in Papers I and II.
For the Sagittarius clusters Terzan 7, Terzan 8, and Arp 2 however,
the annulus widths had to be adjusted. These three clusters are
intrinsically sparse (see Table \ref{params}), and combined with
their relatively close distance ($\sim 25$ kpc) and the small WFPC2 
field of view, this means that the stellar density in the observed
frames is very low. These three clusters therefore required significantly
wider annuli to obtain useful profiles. A small amount of experimentation
resulted in the following widths: 
Terzan 7 -- $2\arcsec$, $3\arcsec$, $4\arcsec$, and $6\arcsec$; 
Terzan 8 -- $7\arcsec$, $8\arcsec$, $9\arcsec$, and $10\arcsec$; 
Arp 2 -- $8\arcsec$, $9\arcsec$, $10\arcsec$, and $11\arcsec$.

We calculated the surface brightness $\mu_i$ of the $i$-th annulus
in a given annulus set by summing the flux of all stars contained in
that annulus:
\begin{equation}
\mu_{i} = \frac{A_{i}}{\pi (b^{2}_{i} - a^{2}_{i})} \sum_{j=1}^{N_{s}} C_{j} F_{j}   
\label{sb}
\end{equation}
where $a_{i}$ and $b_{i}$ are respectively the inner and outer radii 
of the annulus, $N_{s}$ is the number of stars in the annulus, $F_{j}$
is the flux of the $j$-th star, and the coefficients $A_{i}$ and $C_{j}$
are correction factors for the annulus area and the detection completeness
respectively. The area correction factors arise because most annuli
are not fully ``imaged'' by the WFPC2 field of view, as described in
Paper I. Therefore the effective area 
$\pi A_{i}^{-1} (b^{2}_{i} - a^{2}_{i})$ must be used in calculating the 
surface brightness of the annulus, rather than the full area. Because of 
the complicated geometry of the WFPC2 camera, and the arbitrary cluster
centering and observation roll-angle, the coefficients $A_i$ were
computed numerically rather than analytically. To avoid
large random uncertainties, annuli with $A_i > 4$ were not used --
this determined the maximum extent $r_m$ of each surface brightness 
profile. In general, $r_m \sim 75 \arcsec$.

The completeness correction factors are necessary because in crowded
fields such as those in globular clusters, detection software cannot
find every star all the time. This results in missing flux, which
should be accounted for when calculating the surface brightness
in each annulus. As described in Paper I, we derived the coefficients 
$C_j$ by means of artificial star tests. Fake stars were added by 
{\em multiphot} one at a time to each frame and then solved, with 
the output filtered according to the detection parameters, just as
for the real photometry. By fully sampling an artificial CMD, for a 
given cluster the completeness was generated as a function of position 
on the camera, and stellar brightness and colour, in the form of a 
look-up table.  For the $j$-th star in an annulus the appropriate $C_j$ 
was located in this table, and the completeness corrected stellar flux 
$C_j F_j$ added to the sum. We discarded stars with $C_j > 4$ to
avoid introducing large random errors -- in effect, this determined the
faint limit as a function of position within the cluster.

Because of the use of both short and long exposures in constructing the
Fornax clusters' master frames, none of them suffered significant
saturation or crowding, except in the central-most few arcseconds in 
the compact Fornax clusters 3, 4, and 5. Apart from M54, none of the 
Sagittarius clusters suffered from crowding either -- as discussed 
previously, these three clusters are sparsely populated, especially 
Terzan 8 and Arp 2. However, 
as with the low density old LMC clusters (e.g., NGC 1466, NGC 2210, 
NGC 2257) the three low density Sagittarius clusters had a smattering of 
saturated stars (giant branch and horizontal branch members) which were 
not measured by {\em multiphot} and which did not contribute to their
respective profiles. As discussed in Paper I, we were comfortable in
neglecting these stars from the calculations, because in doing so we
constructed less noisy profiles without compromising the measurement
of structural parameters (see also e.g., Elson et al. \shortcite{efl};
Elson \shortcite{elsonbig}; Elson \shortcite{elsonfive}).

M54 however, is the one of the most luminous local globular clusters 
(e.g., Sarajedini \& Layden \shortcite{sarlaya}) and is very compact.
The images of this cluster therefore suffered from severe saturation
and crowding over much of the PC chip, on which M54 was centered. 
In Paper I, we described our technique of adding photometry from a
short exposure to alleviate such problems; however, unfortunately no
such short exposures are available in the {\em HST} archive for M54.
This meant we were unable to extend the profile for this cluster
within approximately $7\arcsec$. As with the compact old clusters
in the LMC bar (e.g., NGC 1754, NGC 1786, NGC 2005, etc) there were
also a number of saturated (giant) stars outside the central region of 
the cluster -- again, as described above, it was acceptable to 
neglect these stars from the profile calculations.

For the three Fornax clusters which were saturated and crowded within
$\sim 5\arcsec$ radius, we were able to use photometry from individual 
aligned short exposures (see Table \ref{fornaxdata}) from the respective 
data-groups to construct complete profiles, exactly following the
procedure described in Paper I.

For every annulus, the uncertainty $\sigma_i$ in the surface brightness
was initially calculated by dividing the annulus into eight segments,
and evaluating the standard deviation of the surface brightnesses of
the segments. This technique proved unsuitable for the outer regions
of many clusters however, significantly underestimating the 
point-to-point scatter. To account for this, after the background
subtraction (see below) on a profile, we calculated the Poisson
error for each annulus, and in cases where this was significantly larger
than the segmental error, we adopted the new uncertainty in
preference. As a consequence of this technique, for some profiles the
errors appear larger than the RMS point-to-point scatter. For a full
discussion of this, and the error calculation procedure in general,
we refer the interested reader to Paper I.

We next fit models to each measured profile. For Magellanic clusters, 
we found the most suitable models to be those of Elson, Fall \&
Freeman \shortcite{eff} (hereafter EFF models), which have the form:
\begin{equation}
\mu(r) = \mu_{0} \left(1 + \frac{r^{2}}{a^{2}} \right)^{-\frac{\gamma}{2}}
\label{ep}
\end{equation}
where $\mu_{0}$ is the central surface brightness, $a$ is a measure of 
the scale length and $\gamma$ is the power-law slope at large radii.
This is essentially an empirical King \shortcite{king} model, without
the tidal truncation -- because of the WFPC2 field of view, no profiles
of Magellanic Cloud clusters were ever measured to their tidal radii. 
Clearly, for the Sagittarius clusters (much closer than the Magellanic
Cloud clusters) such models were again sufficient. For the Fornax
clusters however, previous measurements (see e.g., Webbink 
\shortcite{webbink}; Demers, Kunkel \& Grondin \shortcite{demersa};
Demers, Irwin \& Kunkel \shortcite{demersb}; Rodgers \& Roberts 
\shortcite{rr}; Smith et al. \shortcite{smithae}) suggested
tidal radii in the range $\sim 50\arcsec - 120 \arcsec$, so it seemed
possible that we {\em would} measure to the tidal limits of some
clusters. Ultimately however, we did not observe any apparent cut-offs
(see Section \ref{sparr}) and EFF profiles again proved sufficient.

The traditional King core radius $r_c$ is related to the EFF scale
length $a$ by the relation:
\begin{equation}
r_{c} = a(2^{2 / \gamma} - 1)^{1 / 2}
\label{rc}
\end{equation}
provided the tidal cut-off $r_{t} >> r_{c}$ -- a safe assumption for
all the clusters measured here.

The best fitting EFF profiles were determined via weighted least-squares
minimization; again for a full discussion, we refer the reader to
Paper I. As part of the fitting process, we estimated a background
level (due to field stars) for each cluster. Because separate offset
background images were not available, this was achieved by first
fitting a model only to the very centre of a given cluster profile
(bright enough to be effectively free of background contamination), 
to obtain estimates for $\mu_0$ and $r_c$. These were then used to
fit a model of the form $r^{-\gamma}+\phi$ to the outer part of the
profile, where $\phi$ is the (constant) background level. We then
subtracted this level from the whole profile, calculated the Poisson
error for each annulus (as described above), and fit an EFF model
to the entire background subtracted profile to obtain final measures
of $(\mu_0,\gamma,a)$. Uncertainties in these parameters were
determined via a bootstrap algorithm, with 1000 iterations (see e.g.,
Press et al. \shortcite{press}, p. 691). We have previously verified that 
this subtraction and fitting procedure does not introduce any large 
systematic errors into the measured values of $\gamma$ -- 
the parameter most sensitive to the assumed background contamination.

Finally, we note that while profiles and parameter measurements are
presented here for only the F555W photometry\footnote{The reason for
this is mainly to do with clarity -- presenting a second full set of
results would tend to unnecessarily extend and dilute the flow of the 
paper.}, similar 
profiles and measurements exist for the F814W photometry (much as 
secondary profiles exist in F450W or F814W for the LMC and SMC clusters 
in Papers I and II). These measurements provide a good consistency
check on the F555W profiles and parameters and also offer the opportunity
for studies of colour profiles, for example. The second colour 
measurements {\em are} available on-line, as described in Section 1.

\section{Results}
\label{results}
\subsection{Profiles and structural parameters}
\label{spar}
\begin{figure*}
\begin{minipage}{175mm}
\includegraphics[width=87mm,height=79.5mm]{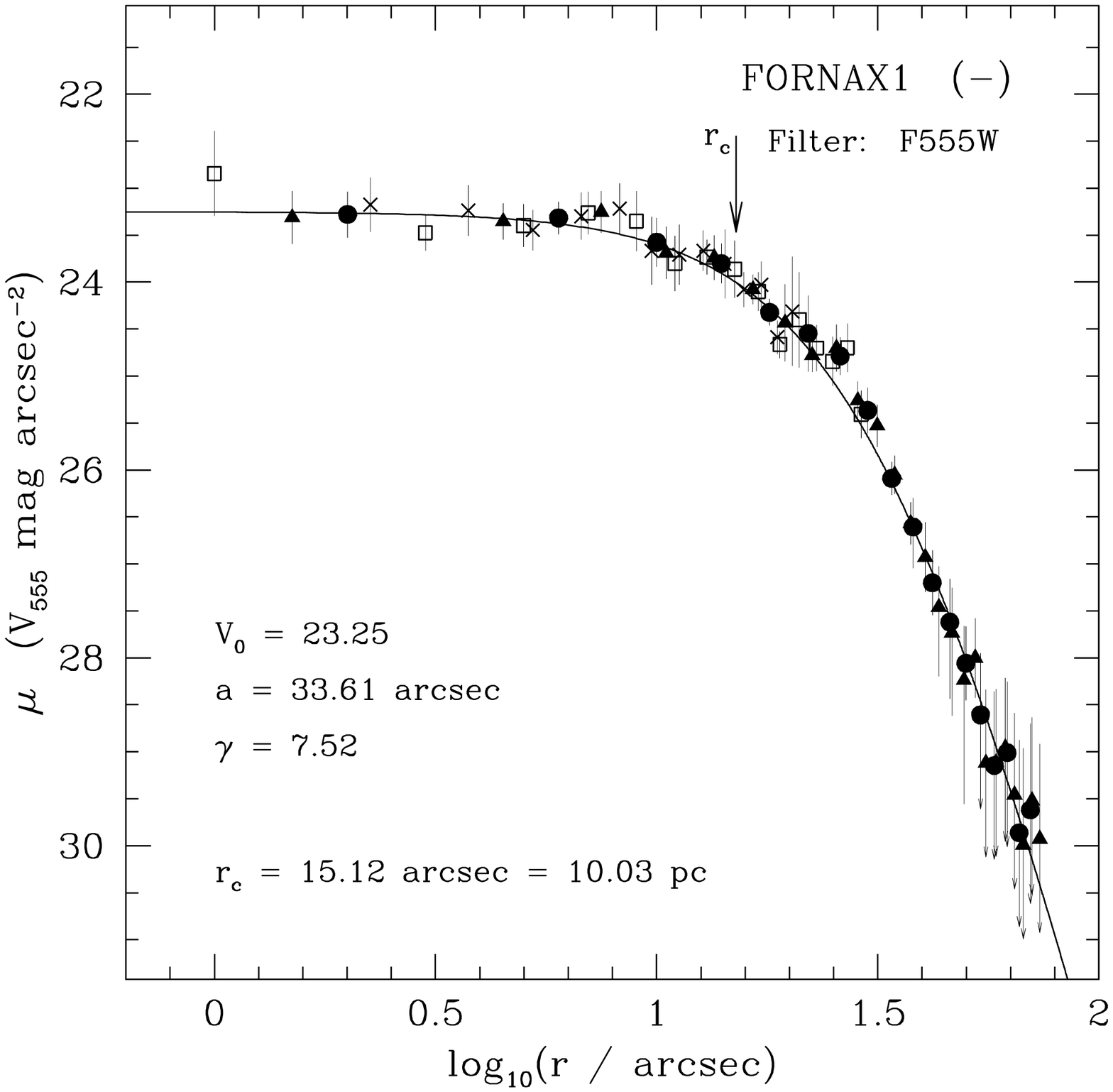}
\includegraphics[width=87mm,height=79.5mm]{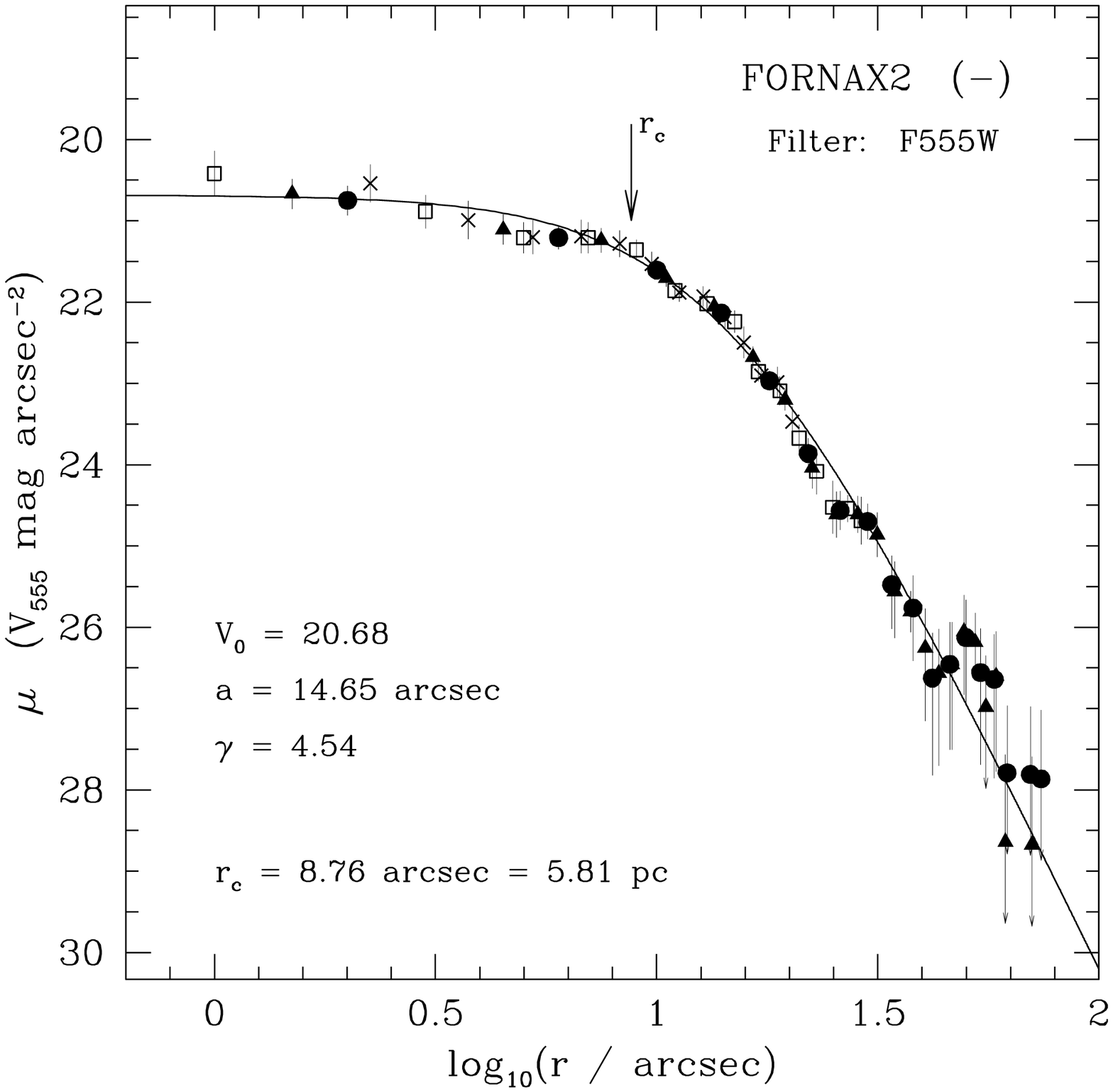}
\newline
\includegraphics[width=87mm,height=79.5mm]{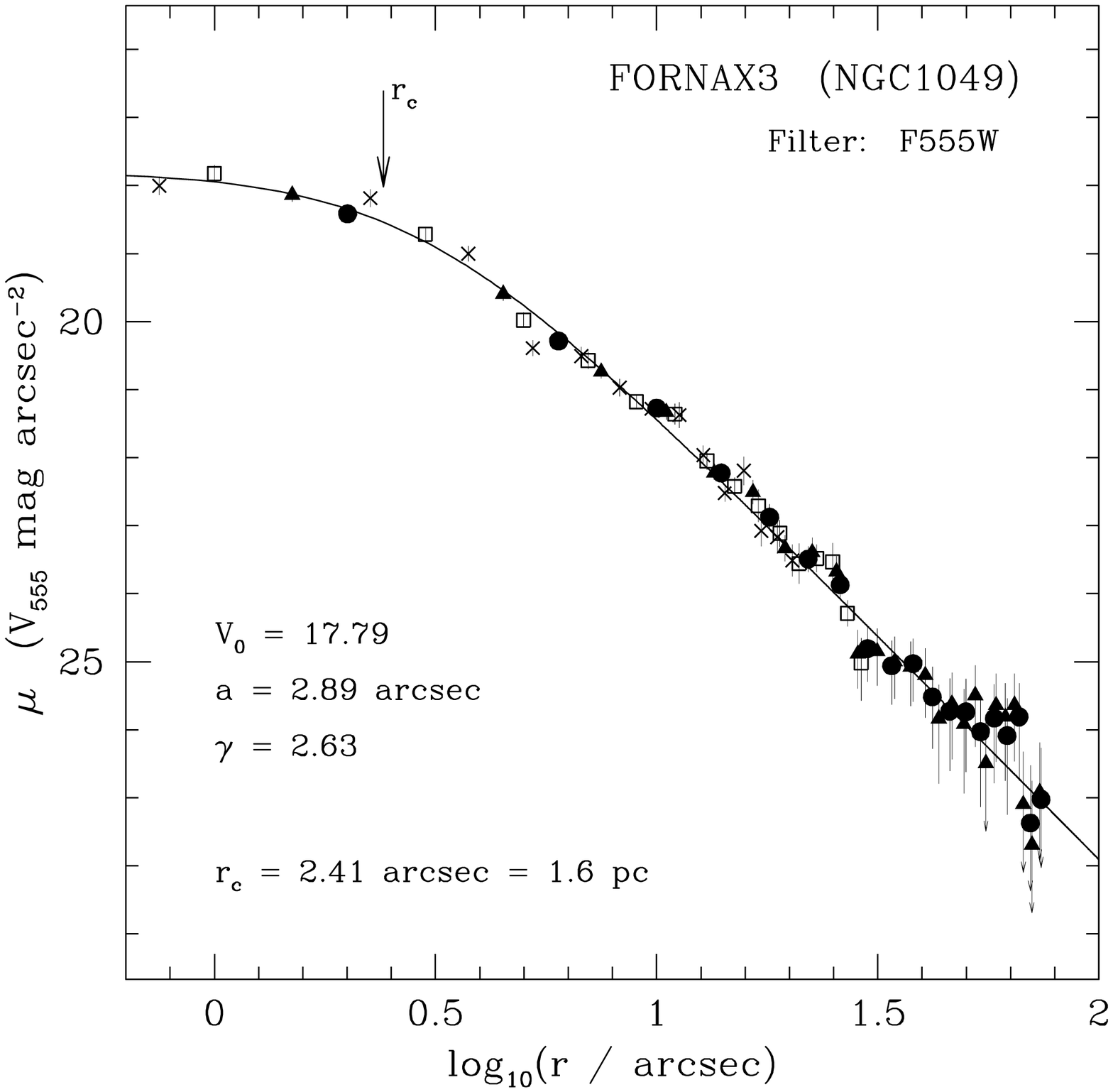}
\includegraphics[width=87mm,height=79.5mm]{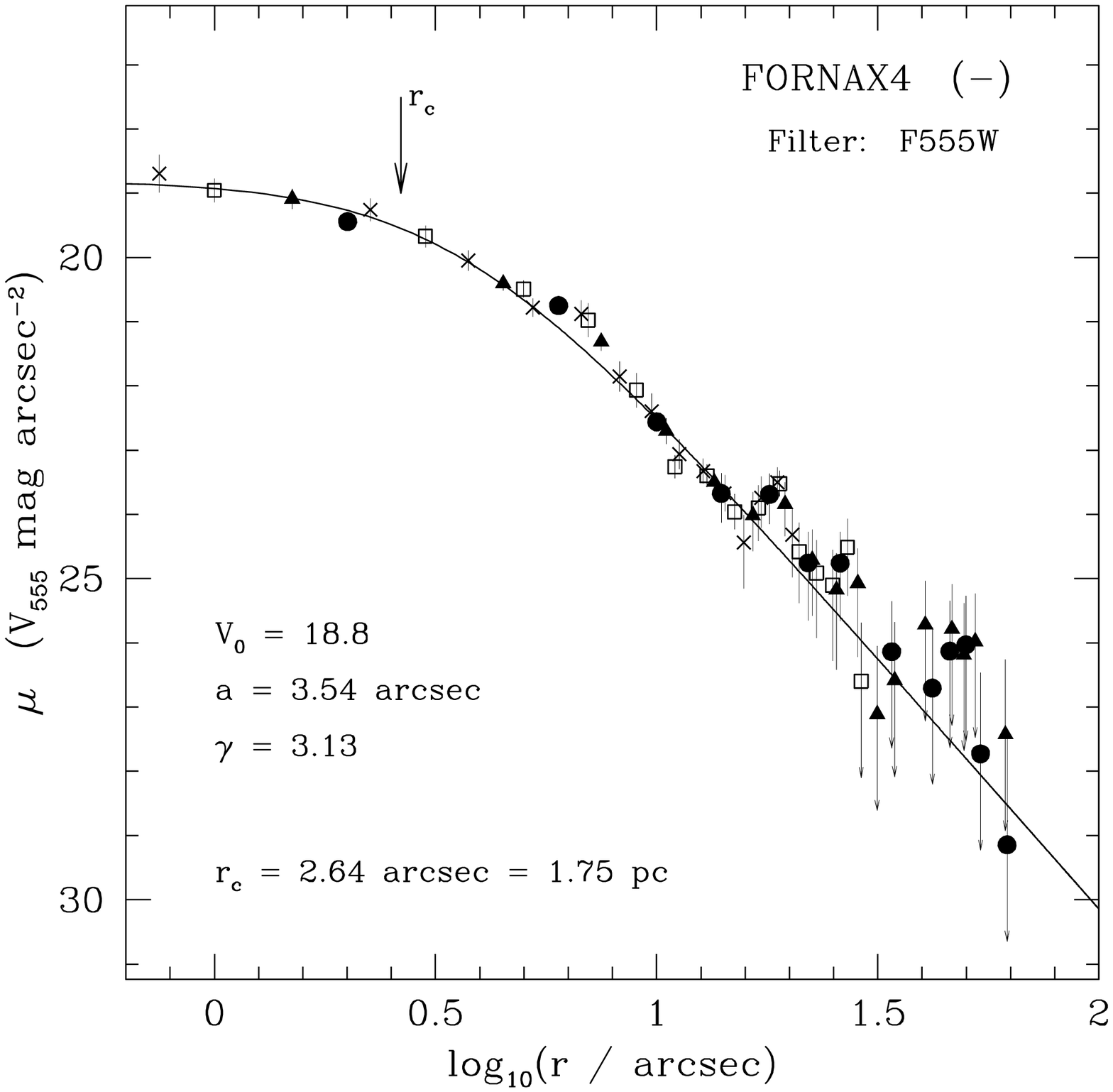}
\newline
\includegraphics[width=87mm,height=79.5mm]{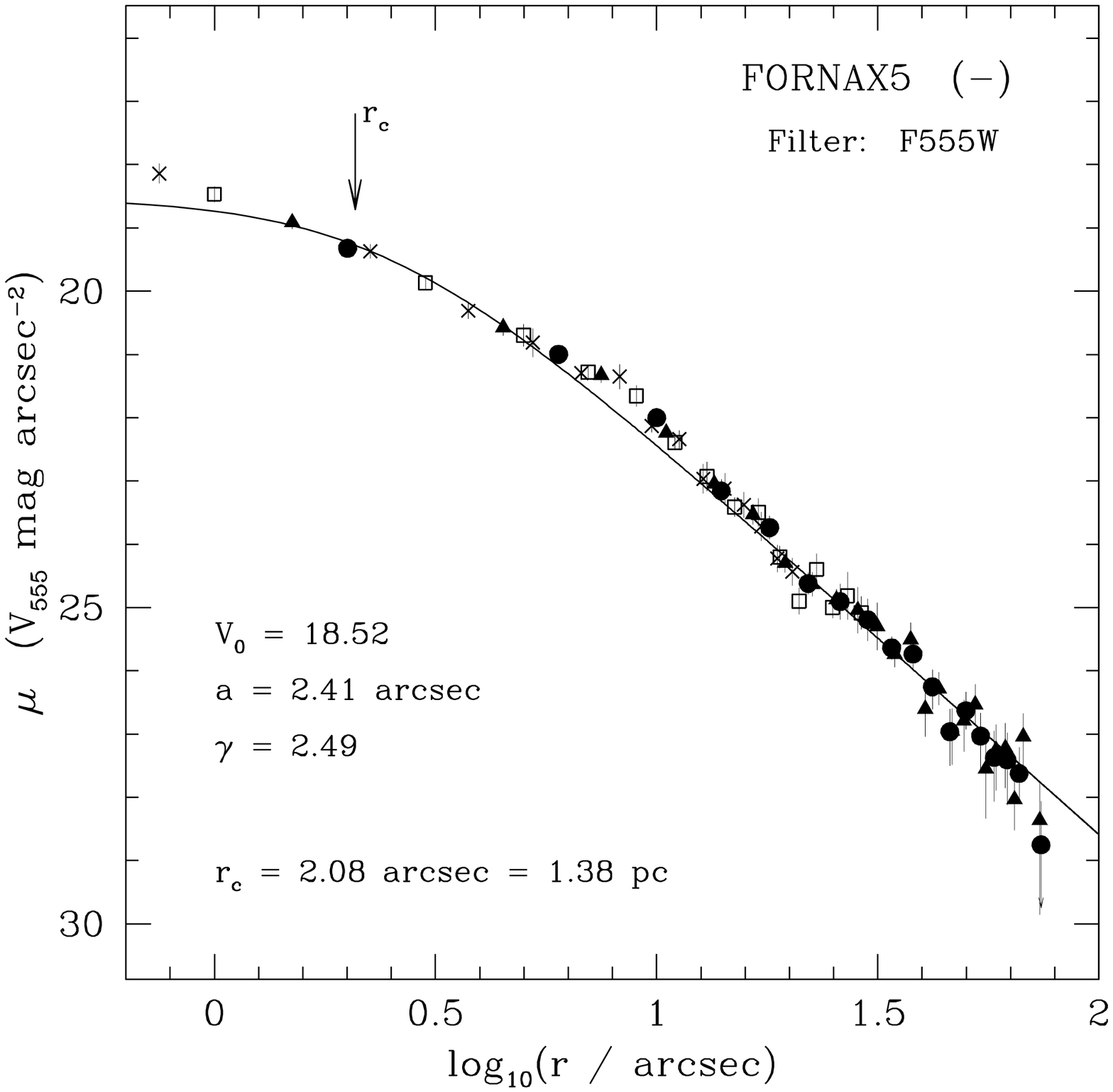}
\hspace{0.5cm}
\begin{minipage}{75mm}
\vspace{-5cm}
\caption{Background-subtracted F555W surface brightness profiles for each of the five Fornax clusters. The four different annulus widths are marked with different point types: $1\farcs5$ width are crosses, $2\arcsec$ width are open squares, $3\arcsec$ width are filled triangles, and $4\arcsec$ width are filled circles. Error bars marked with down-pointing arrows fall below the bottom of their plot. The solid lines show the best-fit EFF profiles. For each cluster the core radius $r_{c}$ is indicated and the best-fit parameters listed. When converting to parsecs, we assume a distance modulus of 20.68 (see text).}
\label{fornaxplots}
\end{minipage}
\end{minipage}
\end{figure*}

\begin{figure*}
\begin{minipage}{175mm}
\includegraphics[width=87mm,height=79.5mm]{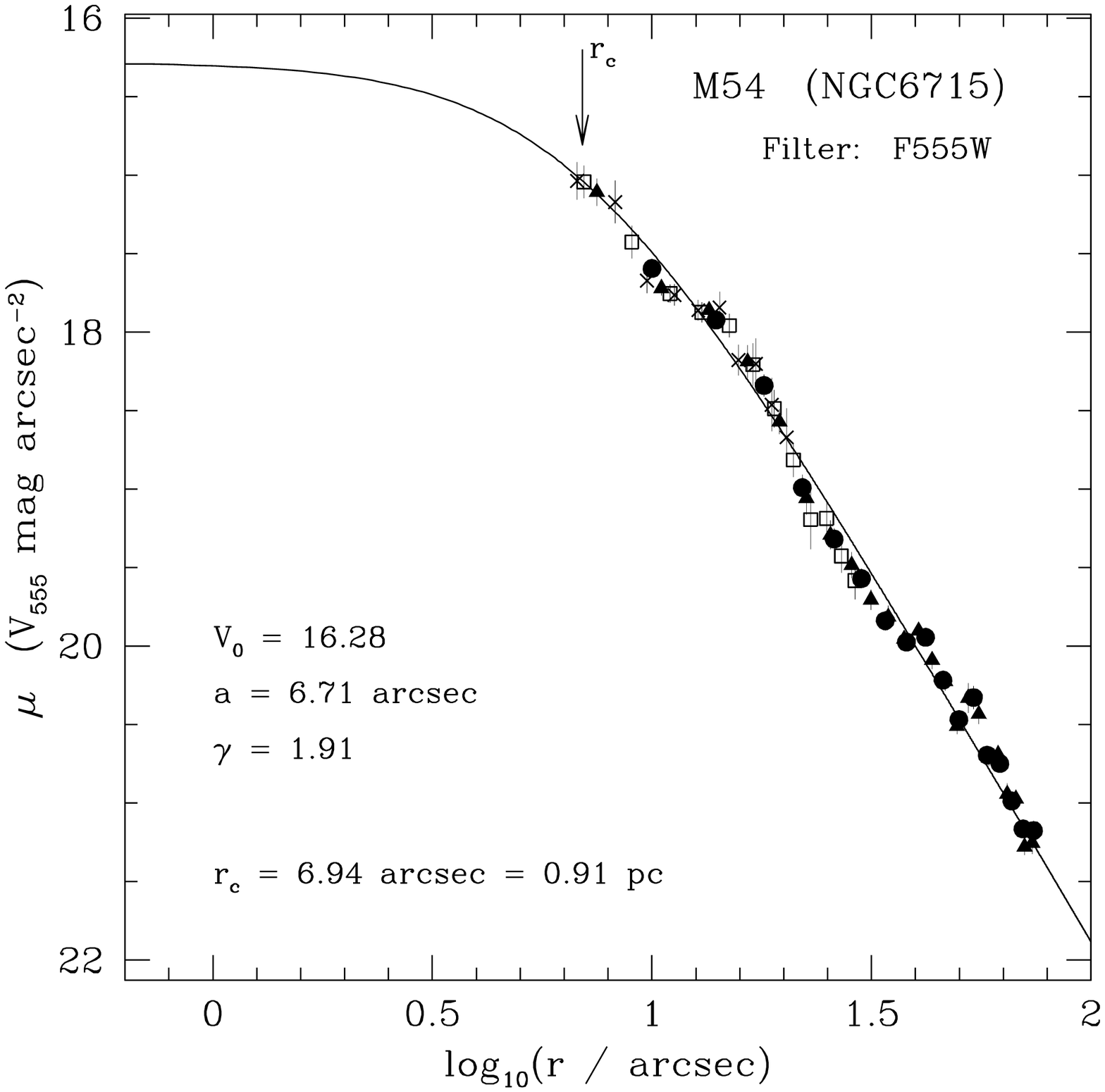}
\includegraphics[width=87mm,height=79.5mm]{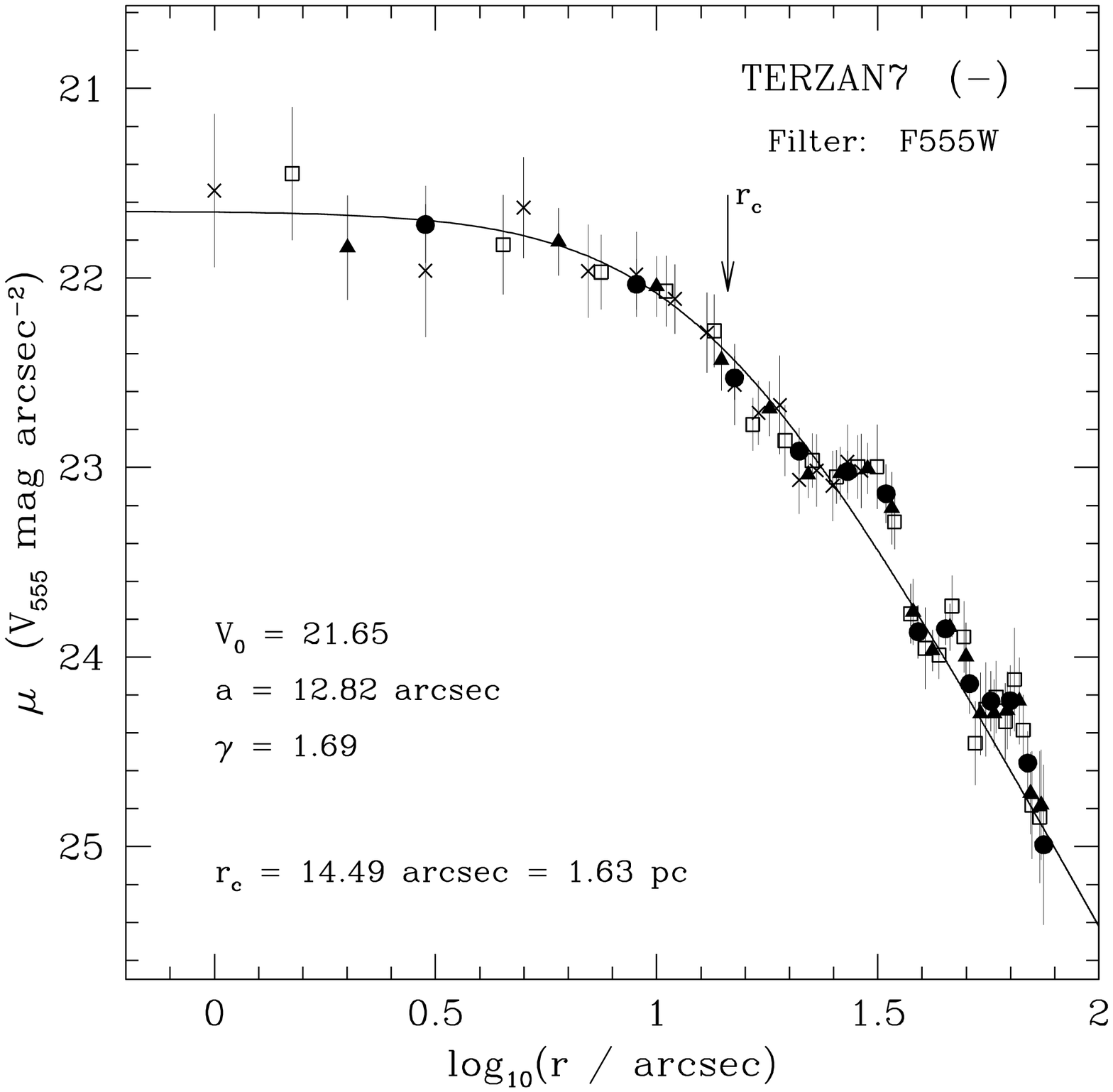}
\newline
\includegraphics[width=87mm,height=79.5mm]{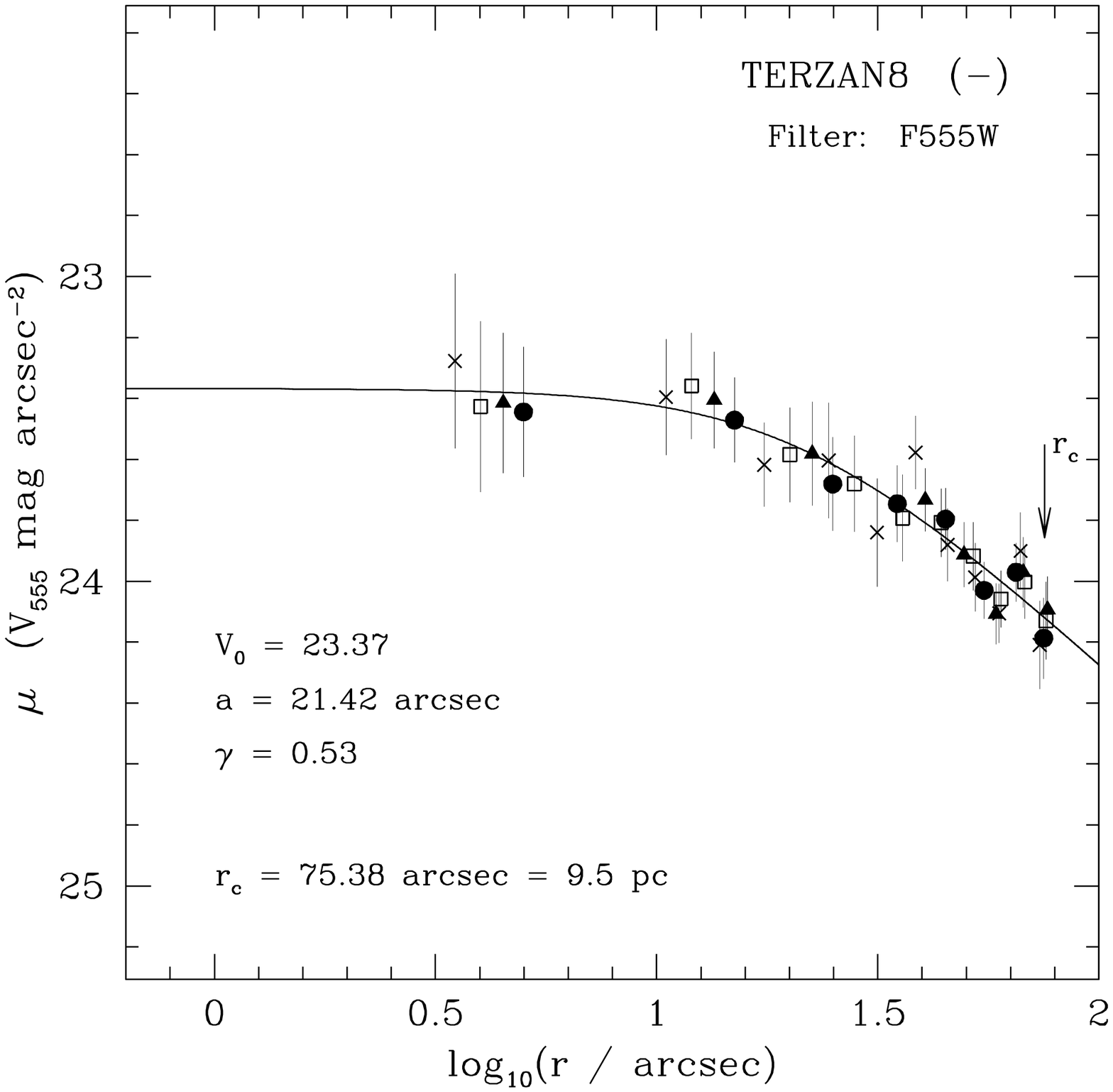}
\includegraphics[width=87mm,height=79.5mm]{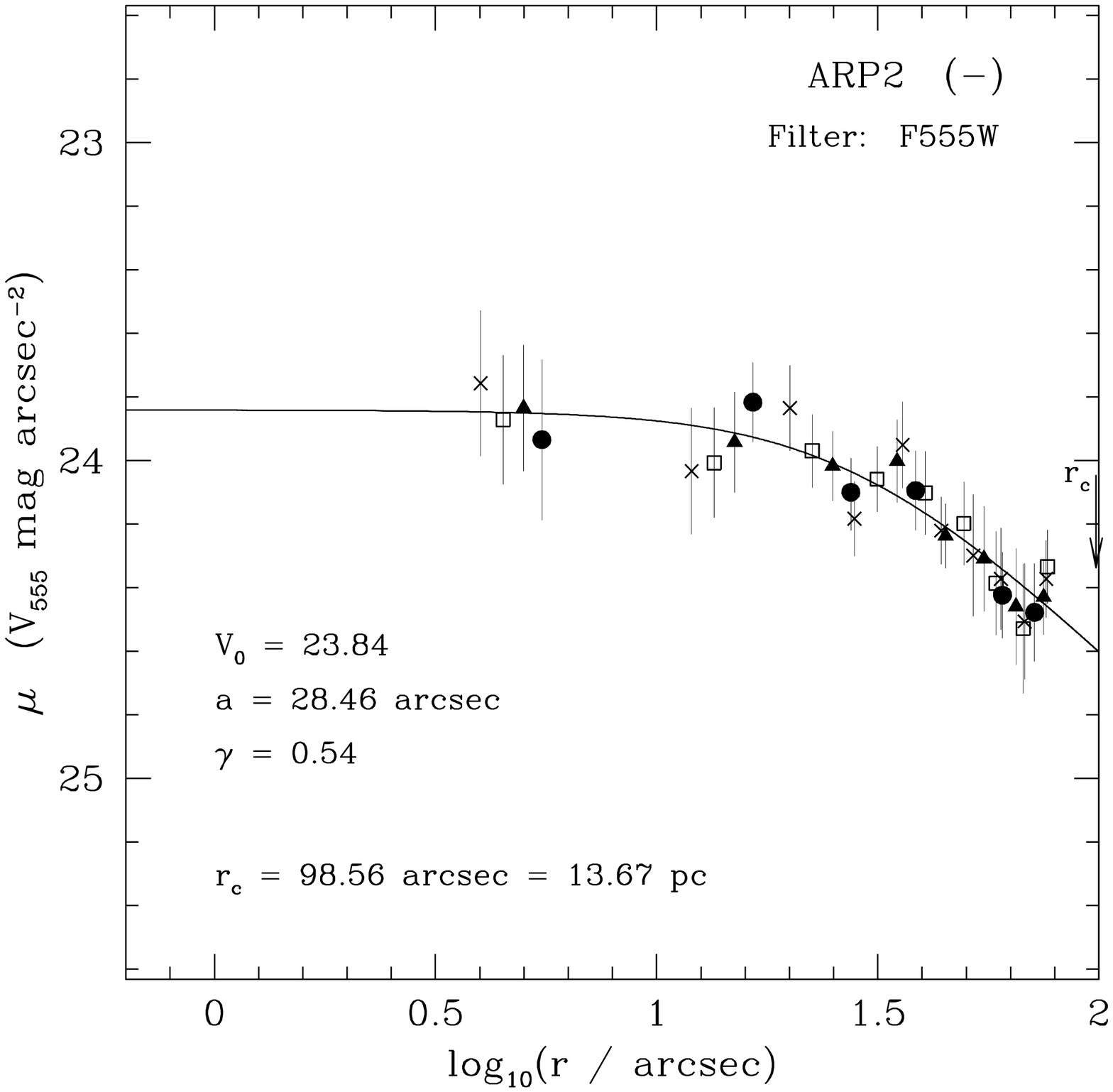}
\caption{Background-subtracted F555W surface brightness profiles for each of the four Sagittarius clusters. For each cluster, the four different annulus widths are marked with different point types. These are (see text): M54: $1\farcs5$ (crosses), $2\arcsec$ (open squares), $3\arcsec$ (filled triangles), $4\arcsec$ (filled circles); Terzan 7: $2\arcsec$ (crosses), $3\arcsec$ (open squares), $4\arcsec$ (filled triangles), $6\arcsec$ (filled circles); Terzan 8: $7\arcsec$ (crosses), $8\arcsec$ (open squares), $9\arcsec$ (filled triangles), $10\arcsec$ (filled circles); Arp 2: $8\arcsec$ (crosses), $9\arcsec$ (open squares), $10\arcsec$ (filled triangles), $11\arcsec$ (filled circles). Error bars marked with down-pointing arrows fall below the bottom of their plot. The solid lines show the best-fit EFF profiles. For each cluster the core radius $r_{c}$ is indicated and the best-fit parameters listed. When converting to parsecs, we assume the distance moduli from Table \ref{distred}.}
\label{sgrplots}
\end{minipage}
\end{figure*}

The background-subtracted F555W surface brightness profiles for the 
Fornax dwarf globular clusters are presented in Fig. \ref{fornaxplots},
and those for the Sagittarius dwarf globular clusters in Fig.
\ref{sgrplots}. The results of the EFF model fitting are listed in
Table \ref{params}. Because of the limiting resolution of our profiles
(set by the smallest annulus width), the measurements of $r_c$ for
the compact clusters Fornax 3, 4, and 5 are best considered as upper
limits.

% SB profiles here... about 2 pages' worth.

\begin{table*}
\begin{minipage}{162mm}
\caption{Structural parameters for the cluster sample derived from the best-fitting F555W EFF profiles.}
\begin{tabular}{@{}lcccccccc}
\hline \hline
Cluster & \multicolumn{2}{c}{Centre (J2000.0)$^{a}$} & $\mu_{555}(0)^{b}$ & $a$ & $\gamma$ & $r_{c}$ & $r_{c}$ & $r_{m}$ \vspace{0.5mm} \\
Name & $\alpha$ & $\delta$ & & $(\arcsec)$ & & $(\arcsec)$ & (pc)$^{c}$ & $(\arcsec)$ \\
\hline
Fornax 1 & $02^{h}37^{m}01\fs9$ & $-34\degr 11\arcmin 01\arcsec$ & $23.25 \pm 0.04$ & $33.61 \pm 2.42$ & $7.52 \pm 0.64$ & $15.12 \pm 0.43$ & $10.03 \pm 0.29$ & $75$ \\
Fornax 2 & $02^{h}38^{m}44\fs1$ & $-34\degr 48\arcmin 30\arcsec$ & $20.68 \pm 0.06$ & $14.65 \pm 0.84$ & $4.54 \pm 0.23$ & $8.76 \pm 0.29$ & $5.81 \pm 0.19$ & $76$ \\
Fornax 3 & $02^{h}39^{m}48\fs1$ & $-34\degr 15\arcmin 30\arcsec$ & $17.79 \pm 0.07$ & $2.89 \pm 0.17$ & $2.63 \pm 0.06$ & $2.41 \pm 0.11$ & $1.60 \pm 0.07$ & $76$ \\
Fornax 4 & $02^{h}40^{m}07\fs6$ & $-34\degr 32\arcmin 10\arcsec$ & $18.80 \pm 0.11$ & $3.54 \pm 0.47$ & $3.13 \pm 0.20$ & $2.64 \pm 0.27$ & $1.75 \pm 0.18$ & $64$ \\
Fornax 5 & $02^{h}42^{m}21\fs1$ & $-34\degr 06\arcmin 07\arcsec$ & $18.52 \pm 0.14$ & $2.41 \pm 0.24$ & $2.49 \pm 0.06$ & $2.08 \pm 0.17$ & $1.38 \pm 0.11$ & $76$ \\
\hline
M54 & $18^{h}55^{m}03\fs3$ & $-30\degr 28\arcmin 46\arcsec$ & $16.28 \pm 0.05$ & $6.71 \pm 0.40$ & $1.91 \pm 0.04$ & $6.94 \pm 0.32$ & $0.91 \pm 0.04$ & $76$ \\
Terzan 7 & $19^{h}17^{m}43\fs9$ & $-34\degr 39\arcmin 29\arcsec$ & $21.65 \pm 0.05$ & $12.82 \pm 1.23$ & $1.69 \pm 0.07$ & $14.49 \pm 1.03$ & $1.63 \pm 0.12$ & $78$ \\
Terzan 8 & $19^{h}41^{m}45\fs2$ & $-34\degr 00\arcmin 03\arcsec$ & $23.37 \pm 0.04$ & $21.42 \pm 6.14$ & $0.53 \pm 0.11$ & $75.38 \pm 5.71$ & $9.50 \pm 0.72$ & $81$ \\
Arp 2 & $19^{h}28^{m}45\fs2$ & $-30\degr 21\arcmin 21\arcsec$ & $23.84 \pm 0.04$ & $28.46 \pm 8.29$ & $0.54 \pm 0.15$ & $98.56 \pm 13.33$ & $13.67 \pm 1.85$ & $81$ \\
\hline
\end{tabular}
\medskip
\\
$^{a}$ We find our centering algorithm to be repeatable to approximately $\pm 1\arcsec$, notwithstanding image header inaccuracies. Given this precision, coordinates in $\delta$ are provided to the nearest arcsecond. Those in $\alpha$ are reported to the nearest tenth of a second, but the reader should bear in mind that at $\delta = -30\degr$, one second of RA corresponds to approximately thirteen seconds of arc -- in other words, the uncertainty in $\alpha$ is slightly smaller than $\pm 0\fs1$.\\
$^{b}$ The $V_{555}$ magnitude of one square arcsecond at the centre of a given cluster.\\
$^{c}$ When converting to parsecs we assume the distance moduli and scale factors from Table \ref{distred}. \\
\label{params}
\end{minipage}
\end{table*}

\begin{table*}
\begin{minipage}{167mm}
\caption{Previously published structural parameter measurements.}
\begin{tabular}{@{}lcccccc}
\hline \hline
Cluster & $r_{c}$ ($\arcsec$) & References & $r_{c}$ ($\arcsec$) & $r_{t}$ ($\arcsec$) & References & $r_m$ ($\arcsec$) \vspace{0.5mm} \\
 & (published)$^{a}$ & & (this paper) & (published)$^{b}$ & & (this paper) \\
\hline
Fornax 1 & $19 \pm 1$; $5.9 \pm 0.6$; $17.6$; $14.4 \pm 4.9$ & $2$; $5$; $6$; $9$ & $15.12 \pm 0.43$ & $56 \pm 5$; $104$; $91 \pm 36$ & $2$; $6$; $9$ & $75$ \\
Fornax 2 & $4 \pm 1$; $5.5 \pm 0.6$; $8.7 \pm 1.0$ & $3$; $5$; $9$ & $8.76 \pm 0.29$ & $74 \pm 5$; $114 \pm 27$ & $3$; $9$ & $76$ \\
Fornax 3 & $4 \pm 1$; $3.4 \pm 0.3$; $1.4 \pm 0.5$ & $3$; $5$; $9$ & $2.41 \pm 0.11$ & $77 \pm 5$; $95 \pm 22$ & $3$; $9$ & $76$ \\
Fornax 4 & $7 \pm 1$; $3.9 \pm 0.4$; $1.0 \pm 0.4$ & $3$; $5$; $9$ & $2.64 \pm 0.27$ & $43 \pm 5$; $66 \pm 15$ & $3$; $9$ & $64$ \\
Fornax 5 & $4 \pm 1$; $3.7 \pm 0.4$; $\sim 2$ (UL); $4.2 \pm 1.6$ & $3$; $5$; $6$; $9$ & $2.08 \pm 0.17$ & $74 \pm 5$; $91$; $76 \pm 18$ & $3$; $6$; $9$ & $76$ \\
\hline 
M54 & $6.3 \pm 1.5$; $6.5 \pm 1.5$; & $1$; $(8,4)$; $9$ & $6.94 \pm 0.32$ & $446 \pm 157$; $446 \pm 215$; & $1$; $(8,4)$; $9$ & $76$ \\
 & $6.3 \pm 0.7$ & & & $445 \pm 104$ & & \\
Terzan 7 & $39.8 \pm 9.2$; $36.3 \pm 17.3$; & $1$; $(8,4)$; $9$ & $14.49 \pm 1.03$ & $199 \pm 71$; $\sim 440$ (OM); & $1$; $(8,4)$; $9$ & $78$ \\
 & $21.8 \pm 8.2$ & & & $209 \pm 84$ & & \\
Terzan 8 & $\sim 60$ (OM); $46.6 \pm 17.6$ & $(7,4)$; $9$ & $75.38 \pm 5.71$ & $\sim 240$ (OM); $523 \pm 210$ & $(7,4)$; $9$ & $81$ \\
Arp 2 & $100 \pm 23.2$; $95.5 \pm 22.2$; & $1$; $(8,4)$; $9$ & $98.56 \pm 13.33$ & $1000 \pm 353$; $\sim 760$ (OM); & $1$; $(8,4)$; $9$ & $81$ \\
 & $119.7 \pm 33.5$ & & & $337 \pm 136$ & & \\
\hline
\end{tabular}
\medskip
\\
Reference list: 1. Chernoff \& Djorgovski \shortcite{chernoff}; 2. Demers et al. \shortcite{demersa}; 3. Demers et al. \shortcite{demersb}; 4. Harris \shortcite{harris} (1999 update); 5. Rodgers \& Roberts \shortcite{rr}; 6. Smith et al. \shortcite{smithae}; 7. Trager et al. \shortcite{tragera}; 8. Trager et al. \shortcite{tragerb}; 9. Webbink \shortcite{webbink}. \\
$^{a}$ Notes regarding errors in $r_c$: Errors are exactly as quoted by the authors for references 2, 3, \& 5. For reference 1, the authors suggest errors of $\sim 0.1$ in $\log r_c$. For reference 6, no errors are quoted but the authors note that $r_c$ for Fornax 5 is an upper limit -- denoted (UL) above. For reference 8, no formal errors are quoted; however in reference 7, which contains mostly the same measurements, data quality flags are indicated. The authors state that clusters with data quality 1 (M54, Arp 2) have errors of $\sim 0.1$ in $\log r_c$, those with quality 2 (Terzan 7) have errors of twice this, and those with quality 4 (Terzan 8) are order of magnitude estimates -- denoted (OM) above. We have used these errors to estimate the uncertainties as listed above. In reference 9, the author states that clusters with $r_c$ from surface photometry (M54, Fornax 2) have errors of $0.049$ in $\log r_c$ ($r_c$ in arcmin), those from star counts (Arp 2) have errors of $\sim 0.12$, and those estimated by eye (Fornax 1, 3, 4, 5, Terzan 7, 8) have errors of $\sim 0.16$. \\
$^{b}$ Notes regarding errors in $r_t$: Errors are exactly as quoted by the authors for references 2 \& 3. For reference 1, the authors suggest errors of $0.1-0.2$ (we assume $0.15$) in $c = \log(r_t/r_c)$, which we use to estimate $r_t$ and its uncertainty. For reference 6, no errors are quoted. For reference 8, again no errors are quoted; however in reference 7, which contains mostly the same measurements, data quality is indicated. The authors state that clusters with data quality 1 (M54) have errors of $\sim 0.2$ in $c$. For clusters with poorer data (Terzan 7, 8, Arp 2) we consider the estimates for $r_t$ as order of magnitude (OM) only. In reference 9, the author states that clusters with $r_t$ determined from star counts (M54) or aperture photometry (Fornax 2, 3, 4, 5) have errors of $\sim 0.1$ in $\log r_t$, and those estimated from angular diameter measurements (Fornax 1, Terzan 7, 8, Arp 2) have errors of $\sim 0.17$.  \\
\label{prevdata}
\end{minipage}
\end{table*}

\begin{figure*}
\begin{minipage}{175mm}
\includegraphics[width=87mm]{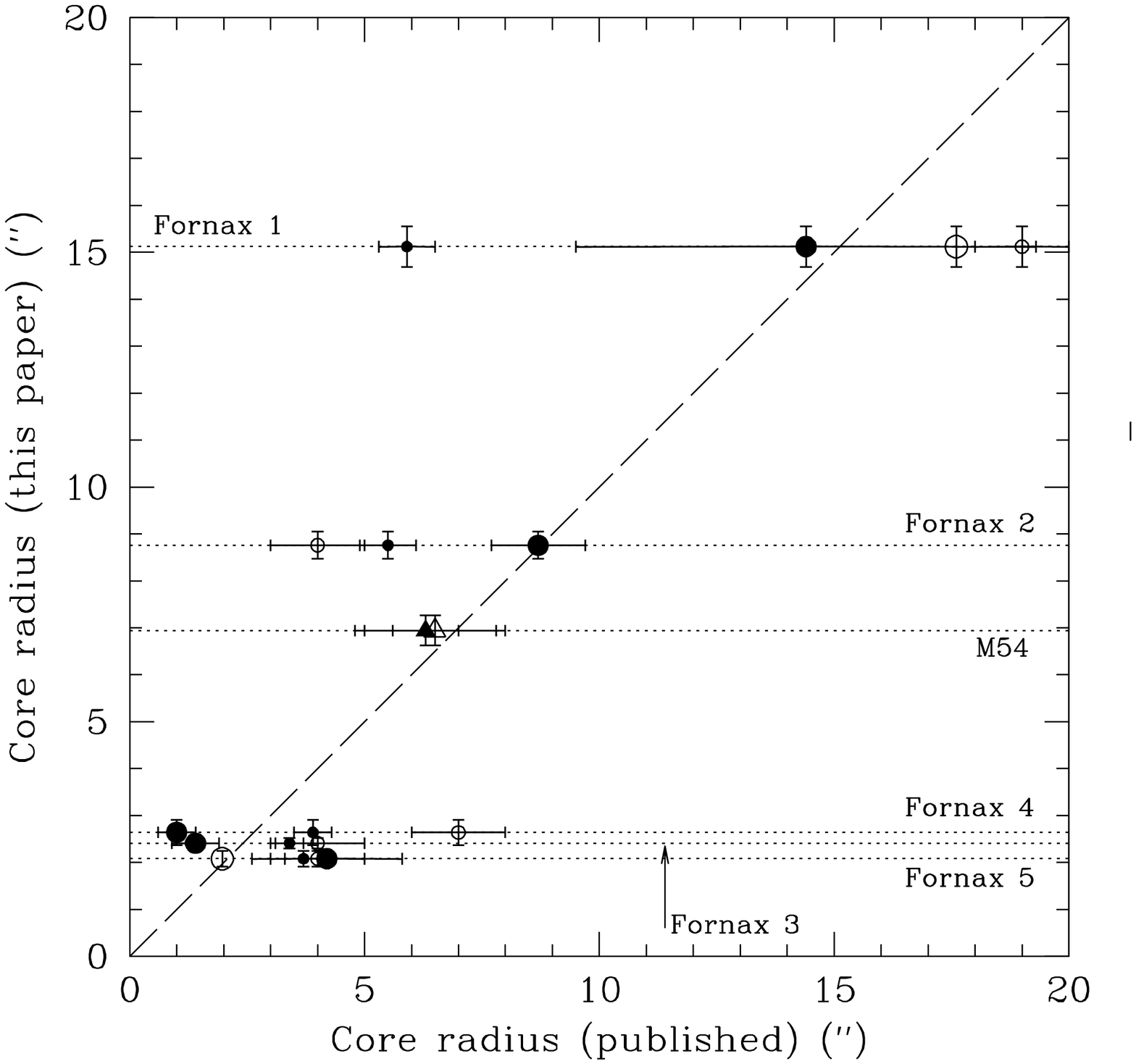}
%\hspace{2mm}
\includegraphics[width=87mm]{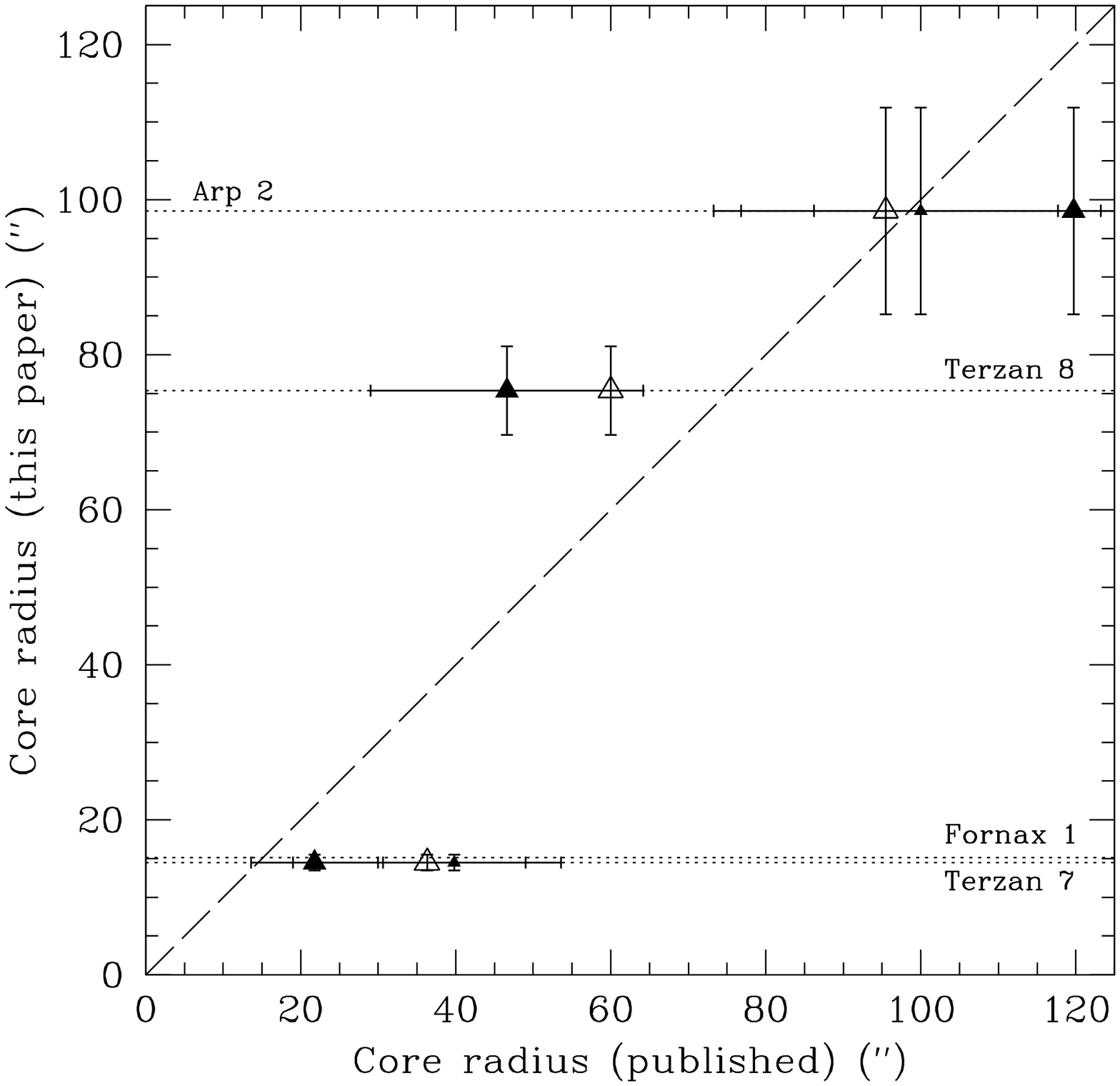}
\caption{Measured values of $r_c$ plotted against those previously published in the literature. {\em Left:} Clusters in the range $0-20\arcsec$. {\em Right:} Expanded to include clusters over the full range. In both plots, the Fornax clusters are represented by circular points and the Sagittarius clusters by triangular points. For the Fornax clusters, the literature references and corresponding point styles are: Demers et al. \shortcite{demersa,demersb} (small open circles); Rodgers \& Roberts \shortcite{rr} (small solid circles); Smith et al. \shortcite{smithae} (large open circles); Webbink \shortcite{webbink} (large solid circles). For the Sagittarius clusters, the literature references and corresponding point styles are: Chernoff \& Djorgovski \shortcite{chernoff} (small solid triangles); Trager et al. \shortcite{tragera,tragerb} and Harris \shortcite{harris} (large open triangles); Webbink \shortcite{webbink} (large solid triangles). On each plot, the dashed line is plotted for reference and represents equality between the measurements. The horizontal dotted lines represent the core radius measurements made in this paper, and allow the set of literature values for any given cluster to be identified. In the full-range plot, data points for the small $r_c$ clusters have been omitted for clarity. The dotted line for Fornax 1 has been retained however, to provide some idea of the scale.}
\label{compareplots}
\end{minipage}
\end{figure*}

\subsubsection{Comparison with previous work}
\label{sparr}
It is worthwhile comparing our results with those from previously 
published studies. The structures of the Fornax clusters have been
measured by Demers, Kunkel \& Grondin \shortcite{demersa}
(Fornax 1); Demers, Irwin \& Kunkel \shortcite{demersb} (Fornax 2, 3,
4, 5); Rodgers \& Roberts \shortcite{rr} (all clusters); and Smith et al.
\shortcite{smithae} (Fornax 1 \& 5). Webbink \shortcite{webbink} also
includes estimates of the structural parameters for all five clusters
in his compilation for Galactic globular clusters. The Sagittarius 
clusters are less well studied -- profiles are presented for M54, 
Terzan 7, and Arp 2 by Trager, King \& Djorgovski \shortcite{tragerb} 
(although the profiles for Terzan 7 and Arp 2 are of low resolution),
and structural measurements by Chernoff \& Djorgovski 
\shortcite{chernoff}. An estimate for the structural parameters of 
Terzan 8 is included in 
the compilation of Trager, Djorgovski \& King \shortcite{tragera}. 
Estimates for all four clusters are also provided in Webbink's 
\shortcite{webbink} collection.

The values for the structural parameters of each cluster, taken from 
the above publications, are displayed in Table \ref{prevdata}, and the 
core radii from this table are plotted against our measured values in 
Fig. \ref{compareplots}. We first consider the Fornax clusters (plotted 
as circles). As is evident from Table \ref{prevdata}, for each of the 
five clusters there is significant scatter in the literature measurements;
however, in many cases the uncertainties are also large. In general,
our agreement with previous measurements is good, and there seem to
be no systematic offsets for any given cluster, or literature data set.
This is consistent with any discrepancies being due to random errors --
especially likely given the difficulties associated with ground-based
measurements of such distant clusters. For example, many of the
literature measurements for the compact clusters Fornax 3, 4, and 5
are larger than the core radii calculated in this paper. This is
consistent with the seeing-limited resolution of ground-based imaging.
The most discrepant measurement is that for the core radius of Fornax 1
from Rodgers \& Roberts \shortcite{rr}; however there is a good 
explanation for this. Fornax 1 has a somewhat patchy appearance, and
the surface brightness profile of Rodgers \& Roberts reflects this
with a large bump at around $10\arcsec$. Their best-fit King model does
not account for this bump, and they therefore measure a smaller value
for $r_c$ than we do here. Their profile however, is also consistent
with a larger $r_c$, similar to that from Demers et al. 
\shortcite{demersa} or Smith et al. \shortcite{smithae}, who also show
very bumpy profiles.

It is also worth considering the tidal radii for the Fornax clusters.
Of the five surface brightness profiles, none shows any good evidence for
a tidal turn-down in its outer regions. A comparison between the 
literature values for $r_t$ and the maximum extent of our surface 
brightness profiles shows that it is indeed unlikely that we have 
measured past the tidal cut-offs of any of the clusters
except perhaps Fornax 4. In this case however, the background level
is quite high and the uncertainties intrinsic to our subtraction 
procedure have removed any fine detail in the outer region of this
profile. We are therefore justified in our decision to fit EFF models
to all our Fornax cluster profiles. Finally, we note that Rodgers \& 
Roberts \shortcite{rr} found that their profiles for Fornax clusters 3, 
4, and 5 did not show any evidence for tidal truncation but instead 
demonstrate that these clusters apparently possess extended haloes. Any 
such haloes are worthy of further attention; unfortunately however, our 
profiles are somewhat too limited in extent to confirm or refute their 
existence.

The profiles for the Sagittarius clusters also require careful 
consideration. In this case, the key question is not whether the
profiles have been measured too far radially, but whether they have been
measured far enough. The profile for M54 is incomplete in its inner
regions, while the clusters Terzan 8 and Arp 2 are so extended that
their profiles are only measured to $\sim r_c$. It is therefore not
clear that we are justified in considering many of our best fit parameters
as good measurements for these clusters. 

In particular, it is evident that for all four of the Sagittarius 
clusters, $\gamma < 2$. This is caused by the fact that we have measured
nowhere near to the tidal limits of these clusters, as shown in Table
\ref{prevdata}. Although theoretically $\gamma$ is the power-law slope
of a profile at $r >> a$, $\gamma$ as determined from a best-fitting EFF
model is instead the power-law slope of the measured profile 
{\em near its maximum radial extent}. Given that none of the Sagittarius 
clusters have been
measured to near their tidal radii, it seems likely that each of the
four profiles drops off much more steeply beyond $r_m$ than our
values of $\gamma$ indicate, especially for Terzan 8 and Arp 2. 
Therefore, unlike for the Fornax clusters, where $\gamma$ is a good 
measure of the true power-law slope at large $r$, we must consider our 
measures of $\gamma$ for the Sagittarius clusters as (very) lower limits 
for this quantity.

It is intriguing however, that severely under-estimating $\gamma$ does
not seem to systematically affect our measurements of $r_c$ for
these clusters. Our value for the core radius of M54 is in good
agreement with the previously published quantities, even though
we are missing the inner portion of the profile and our $\gamma$ is
a lower limit. In addition, our measurement of $\mu_{555}(0)$
for this cluster is likely quite uncertain given that we have no inner
data point to tie it down. Our value for the core radius of Terzan 7
is considerably smaller than the literature values, although is within
the uncertainties for Webbink's \shortcite{webbink} measurement. We do
not find cause for concern in this discrepancy -- for example, 
the $r_c$ of Trager et al. \shortcite{tragerb} is measured from a low 
resolution profile which shows evidence of mis-centering -- likely due 
to the ragged appearance of Terzan 7.

Terzan 8 and Arp 2 are extremely extended clusters with core radii which
barely fit within the WFPC2 field of view. Nonetheless, our measured
$r_c$ for Arp 2 is in excellent agreement with all three previously 
published quantities. This, along with M54, demonstrates that even 
with $\gamma$ severely under-estimated we can obtain useful measures of 
$r_c$, and leads us to have confidence in our core radii for the other 
Sagittarius clusters.
Terzan 8 is interesting because the two previous measurements  of $r_c$
are respectively ``little better than a guess'' \cite{tragerb}, and 
a by-eye estimate \cite{webbink}. Given the success of our fit to Arp 2,
we feel that our new core radius measurement for Terzan 8, significantly
larger than the two previous measurements, is the most reliable 
available value. Finally, for the Sagittarius clusters it is worthwhile 
noting that if we assume our calculated core radii to be accurate 
measurements, then given that $\gamma$ is under-estimated for each of
these clusters, it follows from Eq. \ref{rc} that $a$ must also be
an under-estimate of that value which would be measured for a fully
extended profile. Therefore, like for the $\gamma$ values, our 
measurements of $a$ must also be considered lower limits.

\subsubsection{Is Fornax 5 a post core-collapse cluster?}
In Paper I, we identified several of the old LMC globular clusters as post
core-collapse (PCC) candidates. A PCC cluster is characterized by an 
apparent power-law profile in its innermost region, rather than a
profile with a constant density core (such as a King or EFF profile).
Many also show a distinct break at the transition to the power-law
region. Studies of Galactic PCC clusters (see e.g., Lugger, Cohn \&
Grindlay \shortcite{lugger}; Djorgovski \& King \shortcite{dk}) have
shown typical power-law slopes in the range $0.6 < \beta < 1.0$, assuming
the profiles go as $r^{-\beta}$. Similarly, in Paper I, our two best
candidates (NGC 2005 and NGC 2019) had $\beta \approx 0.75$. We 
also measured typical break radii of $\sim 1.3$ pc.

As is evident in Fig. \ref{fornaxplots} and Fig. \ref{sgrplots}, the
only clusters in the present sample which are compact enough to be
PCC candidates are Fornax 3, 4, and 5, and M54. Our profile for M54 is
incomplete in its inner region, so we cannot observe if it follows a
power-law. Nevertheless, we observe no evidence of any break ($1.3$ pc
$\sim 10\arcsec$), and no previous high-resolution profiles (such
as that from Trager et al. \shortcite{tragerb}) show any evidence for
PCC structure. The Fornax clusters are more interesting, because this
is the first study with sufficient resolution to identify any PCC 
structure -- for the previous ground-based observations, atmospheric blur
would have wiped out the innermost profile details on a scale of 
$\sim 2\arcsec$ or more. Fig. \ref{fornaxplots} shows that Fornax 3
and Fornax 4 are well fit by EFF profiles in their central regions,
and neither shows evidence for a break at the expected radius. Fornax 5
however, does not appear well fit by an EFF model, with the measured
profile showing a tell-tale deviation at its innermost two points. 

\begin{figure}
\includegraphics[width=0.5\textwidth]{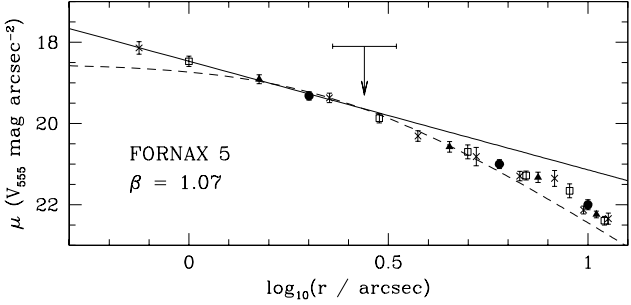}
\vspace{-42mm}
\caption{Power-law fit to the inner core of Fornax 5. The four different point styles represent the four annulus sets, as in Fig. \ref{fornaxplots}. The best-fit EFF and power-law profiles are shown as is the power-law slope ($\beta$) and break radius (arrow) with error indicated. Because of the magnitude scale, the slope of the power-law model as plotted is $2.5\beta$.}
\label{Fornax5core}
\end{figure}

A magnified plot of the core of the profile of Fornax 5 is presented
in Fig. \ref{Fornax5core}, together with an attempted power-law fit,
and the best fitting EFF model. At less than $3\arcsec$ the power-law
is clearly the more suitable model. The best-fitting slope is
$\beta = 1.07$, slightly steeper than the range measured for Galactic
PCC clusters. We estimate a break radius of 
$\log r_{break} = 0.44 \pm 0.08$, which translates to
$r_{break} = 1.83 \pm 0.34$ pc -- slightly larger than the typical
break radius we measured for the LMC PCC candidates. Nonetheless, based
on the evidence of Fig. \ref{Fornax5core}, we believe that Fornax 5
is a solid PCC candidate, worthy of further attention. Interestingly,
in Paper I, we estimated that $20 \pm 7$ per cent of the old LMC cluster
population was in a PCC state, and similarly, Djorgovski \& King 
\shortcite{dk} estimated a PCC fraction of $\sim 20$ per cent for the
Galactic globular clusters. Based on these numbers, it might be expected
that $\sim 1$ cluster in either (or both) of the Fornax and Sagittarius
systems be a PCC cluster -- seemingly exactly what we have observed.

\subsection{Luminosity and mass estimates}
\label{lumandmass}
\begin{table*}
\begin{minipage}{175mm}
\caption{Luminosity and mass estimates calculated using the structural parameters from the best fitting EFF profiles.}
\begin{tabular}{@{}lccccccccc}
\hline \hline
Cluster & $\log \mu_{0}$$^{a}$ & Adopted & Adopted & $\log j_{0}$ & $\log L_{\infty}$ & $\log L_{m}$ & $\log \rho_{0}$ & $\log M_{\infty}$ & $\log M_{m}$ \vspace{0.5mm} \\
 & $(L_{\odot}\,{\rmn pc}^{-2})^{b}$ & [Fe/H] & $M/L_V$ & $(L_{\odot}\,{\rmn pc}^{-3})^{b}$ & $(L_{\odot})$$^{b,c}$ & $(L_{\odot})^{b}$ & $(M_{\odot}\,{\rmn pc}^{-3})$ & $(M_{\odot})^{c}$ & $(M_{\odot})$ \\
\hline
Fornax 1 & $1.32 \pm 0.02$ & $-2.25$ & $3.16$ & $0.00 \pm 0.07$ & $4.07 \pm 0.13$ & $4.07 \pm 0.13$ & $0.50 \pm 0.07$ & $4.57 \pm 0.13$ & $4.57 \pm 0.13$ \vspace{0.2mm} \\
Fornax 2 & $2.39 \pm 0.03$ & $-1.65$ & $3.20$ & $1.31 \pm 0.07$ & $4.76 \pm 0.12$ & $4.75 \pm 0.12$ & $1.81 \pm 0.07$ & $5.26 \pm 0.12$ & $5.26 \pm 0.12$ \vspace{0.2mm} \\
Fornax 3 & $3.50 \pm 0.03$ & $-2.25$ & $3.16$ & $2.99 \pm 0.06$ & $5.06 \pm 0.12$ & $5.00 \pm 0.11$ & $3.49 \pm 0.06$ & $5.56 \pm 0.12$ & $5.50 \pm 0.11$ \vspace{0.2mm} \\
Fornax 4 & $3.20 \pm 0.05$ & $-1.65$ & $2.69$ & $2.64^{+0.13}_{-0.12}$ & $4.69 \pm 0.24$ & $4.67^{+0.23}_{-0.24}$ & $3.07^{+0.13}_{-0.12}$ & $5.12 \pm 0.24$ & $5.10^{+0.23}_{-0.24}$ \vspace{0.2mm} \\
Fornax 5 & $3.24 \pm 0.06$ & $-2.25$ & $3.16$ & $2.79 \pm 0.11$ & $4.76 \pm 0.20$ & $4.67^{+0.17}_{-0.18}$ & $3.29 \pm 0.11$ & $5.25 \pm 0.20$ & $5.17^{+0.17}_{-0.18}$ \vspace{0.2mm} \\
\hline
M54 & $4.23 \pm 0.02$ & $-1.65$ & $3.22$ & $3.97 \pm 0.05$ & $(5.70)$ & $5.36 \pm 0.08$ & $4.48 \pm 0.05$ & $(6.21)$ & $5.86 \pm 0.08$ \vspace{0.2mm} \\
Terzan 7 & $2.00 \pm 0.02$ & $-0.64$ & $2.54$ & $1.49 \pm 0.07$ & $(3.68)$ & $3.50^{+0.10}_{-0.11}$ & $1.90 \pm 0.07$ & $(4.08)$ & $3.91^{+0.10}_{-0.11}$ \vspace{0.2mm} \\
Terzan 8 & $1.37 \pm 0.02$ & $-2.25$ & $3.14$ & $0.24^{+0.23}_{-0.21}$ & $(4.08)$ & $3.67^{+0.10}_{-0.14}$ & $0.74^{+0.23}_{-0.21}$ & $(4.57)$ & $4.17^{+0.10}_{-0.14}$ \vspace{0.2mm} \\
Arp 2 & $1.16 \pm 0.02$ & $-1.65$ & $2.85$ & $-0.13^{+0.25}_{-0.24}$ & $(4.18)$ & $3.59^{+0.09}_{-0.14}$ & $0.33^{+0.25}_{-0.24}$ & $(4.64)$ & $4.04^{+0.09}_{-0.14}$ \vspace{0.2mm} \\
\hline
\end{tabular}
\medskip
\\
$^{a}$ Corrected for reddening using the values from Table \ref{distred}. \\
$^{b}$ Parameters with units of $L_{\odot}$ are F555W luminosities (or luminosity densities). \\ 
$^{c}$ Values in parenthesis (the Sagittarius clusters) are calculated via a scaling estimate, as described in the text. \\
\label{luminmass}
\end{minipage}
\end{table*}

It is possible to estimate luminosities and masses for each cluster
using the structural parameters obtained from the surface brightness
profiles. The procedure for doing this, and the derivation of the
equations involved, is detailed in Paper I. Briefly, we determined
that for a cluster with an EFF surface brightness profile, 
the asymptotic luminosity is given by:
\begin{equation}
L_{\infty} = \frac{2 \pi \mu_{0} a^{2}}{\gamma - 2}
\label{linf}
\end{equation}
provided $\gamma > 2$. When $\gamma \approx 2$, the extrapolation
$r\rightarrow\infty$ likely severely overestimates $L_{\infty}$. 
It is therefore useful to derive a lower-limit for the total luminosity 
of a cluster. We decided that the enclosed luminosity within a cylinder 
of radius $r_{m}$ along the line of sight, where $r_m$ is the maximum 
radial extent measured for a given profile (see Table \ref{params}),
was a suitable quantity, because this is essentially the directly
observed luminosity of a given cluster. In Paper I, we showed that:
\begin{equation}
L_m = \frac{2 \pi \mu_{0}}{\gamma - 2} \left( a^2 - a^\gamma (a^2 + r_m^2 )^{-\frac{(\gamma - 2)}{2}} \right)
\label{lrmax}
\end{equation}
To calculate $L_{\infty}$ and $L_m$, the central surface brightness 
$\mu_{0}$ must be converted to physical units 
($L_{\odot}\,{\rmn pc}^{-2}$). Again in Paper I, we derived an expression
for this:
\[
\log \mu_{0} = 0.4(V_{555}^{\odot} - \mu_{555}(0) + DM + A_V)
\]
\begin{equation}
\hspace{45mm} + \log(SF^{2})\,\, L_{\odot} \,{\rmn pc}^{-2}
\label{convlsun}
\end{equation}
where $DM$ is the distance modulus of the cluster concerned (Table 
\ref{distred}, column 2), $SF$ is its scale factor (arcsec pc$^{-1}$; 
Table \ref{distred}, column 5), $A_V$ is the line of sight $V$-band
extinction, and $V_{555}^{\odot} = +4.85$ is the F555W magnitude of
the sun (see Paper I). To calculate $A_V$ for each cluster, we use
the colour excesses from Table \ref{distred} (either column 6 or 7),
and the extinction laws described in Section \ref{secred}. 

The values for $\mu_0$, $j_0$, $L_{\infty}$, and $L_{m}$ so calculated
appear in Table \ref{luminmass}. We note that these are F555W luminosities
and luminosity densities. As discussed in Section \ref{spar},
all of the Sagittarius clusters have $\gamma < 2$, because only their
very central regions were imaged. This means that $L_{\infty}$ cannot
be directly calculated for these clusters, because the integration which 
gives Eq. \ref{linf} is divergent. However, because the values of
$L_{m}$ are not strictly comparable between clusters or cluster sets
(because $r_m/r_c$ is different for each cluster), we
would like to have some estimates of $L_\infty$. We can obtain these
by using a simple scaling approximation between clusters with
similarly shaped profiles. First, we define a new quantity, $L_c$, which
is simply Eq. \ref{lrmax} with $r_m$ replaced by $r_c$. This represents
the integrated luminosity of a cluster's core. Dividing this by
Eq. \ref{linf} gives:
\begin{equation}
\frac{L_c}{L_\infty} = 1 - a^{(\gamma - 2)} (a^2 + r_c^2 )^{-\frac{(\gamma - 2)}{2}}
\label{corelum}
\end{equation}
For cluster Fornax 1, which has a core radius comparable to Terzan 8 and
Arp 2, this fraction is $\sim 0.4$. For the old clusters with the two
largest core radii from Papers I and II (NGC 1841 and NGC 339) we
calculate $\sim 0.32$ and $\sim 0.35$ respectively. Assuming Terzan 8 
and Arp 2 have profiles which fall off similarly to the profiles of
these three very extended clusters (i.e., $\gamma$ in the range $5-8$)
we can adopt $L_c / L_\infty \sim 0.35$ and estimate $L_\infty$ by
calculating $L_c$. These values are shown in Table \ref{luminmass}. 
Terzan 7 is a more unusual cluster, with a small core radius but 
relatively low central density. Its profile seems most similar to the 
intermediate age clusters NGC 2193, NGC 2213, and Hodge 14, in the LMC. 
These three clusters typically have $L_c / L_\infty \sim 0.12$.
Finally, M54 seems most similar to the compact old bar clusters from the 
LMC study. These clusters have an average $L_c / L_\infty \sim 0.15$.
However, using this to estimate $L_\infty$ for M54 yields a value
smaller than our calculated $L_m$. This suggests that the asymptotic
$\gamma$ for M54 is shallower that those ($\gamma \sim 2.6$) for the
LMC bar clusters, which makes sense seeing as there is probably a much
stronger tidal field at the centre of the LMC than at the centre of
the Sagittarius dwarf. We therefore calculate an upper limit for the 
mass of M54 by extrapolating its profile to the literature tidal radius
from Table \ref{prevdata} and integrating via Eq. \ref{lrmax}. 

The $L_\infty$ measurements agree well with previous such measurements
in the literature. For example, Webbink \shortcite{webbink} lists the
five Fornax clusters as having absolute integrated $V$ magnitudes of
$M_V =$ -5.23, -7.30, -8.19, -7.23, and -7.38, respectively. These
correspond to integrated luminosities of $\log L_{tot} =$ 4.02, 4.85,
5.20, 4.82, and 4.88, respectively, which compare well with the 
$L_\infty$ values derived in the present study. If anything, there is
the tendency for a slight systematic under-estimate in our $L_\infty$
values. This is possibly due to Webbink's slightly larger adopted
distance (145 kpc), although we note that our technique of determining
brightness profiles through star counts does certainly cause some 
luminosity to be missed. Specifically, the bright and faint limits of
each observation exclude some stars from the profiles, leading to an
under-estimate in $\mu_0$ and hence $L_\infty$. For the Fornax clusters,
the missing number of bright stars is negligible due to the inclusion
of short exposures in the construction of the ``master'' frames. For
the Sagittarius clusters no stars above approximately horizontal branch 
level were counted. At the faint end, typical limiting magnitudes for
the Fornax clusters were $V_{555}\sim 26.5$ (i.e., $L\sim 0.5 L_\odot$),
and for the Sagittarius clusters were $V_{555}\sim 24$ (i.e., 
$L\sim 0.2 L_\odot$). We expect the systematic errors from the missed 
contribution to $L_\infty$ from saturated and unseen stars to be within
the random (and other systematic) uncertainties discussed below.

For the Sagittarius clusters, Harris \shortcite{harris} lists
absolute integrated $V$ magnitudes of $M_V =$ -10.01, -5.05, -5.05, 
and -5.28 (for M54, Terzan 7, Terzan 8, and Arp 2 respectively), which
correspond to integrated luminosities of $\log L_{tot} =$ 5.93, 3.95, 
3.95, and 4.18. Again, the agreement with our derived values of 
$L_\infty$ is close, especially given our estimation technique for these
clusters.

By multiplying the appropriate luminosity values by a suitable 
mass-to-light ratio ($M/L_V$) for the cluster in question, estimates for 
the masses $M_{\infty}$ and $M_{m}$, and the central density $\rho_{0}$
may be obtained. As previously (see Papers I and II), we determined
the $M/L_V$ values from the evolutionary synthesis code of Fioc 
\& Rocca-Volmerange \shortcite{pegase} (PEGASE v2.0, 1999), which 
determines the integrated properties of a synthetic stellar population 
as a function of age. We selected a single burst population, formed
with constant metallicity and according to the IMF of 
Kroupa, Tout \& Gilmore \shortcite{ktg} over the mass range 
$0.1$ to $120 M_{\odot}$. As in Papers I and II, we matched each cluster
to the most suitable of the four available abundances in the evolutionary
synthesis code -- either [Fe/H] $\approx -2.25$, $-1.65$, $-0.64$, 
or $-0.33$, based on the literature metallicities in Table \ref{ages} --
and using the literature age estimates (also in Table \ref{ages}) 
determined $M/L_V$. As demonstrated in Paper I, these calculations are 
relatively insensitive to the chosen metallicity, so we are confident 
in using the estimates in Table \ref{ages} even for clusters (such
as Terzan 7) where there is some discrepancy between photometrically
and spectroscopically determined metallicities. The $M/L_V$ ratios are
however, reasonably sensitive to the adopted age (as can be seen in
the difference between the ratios for Fornax 2 and Fornax 4 -- where
Fornax 4 is assumed to be $\sim$ 3 Gyr younger than Fornax 2), so
uncertainties in the ages potentially translate into significant
systematic errors in the assumed $M/L_V$ ratios.

The total mass and central density values are also shown in Table 
\ref{luminmass}. We estimated $M_{\infty}$ for the Sagittarius clusters 
using the $L_\infty$ values calculated as above.
As in Papers I and II, the indicated errors in Table \ref{luminmass}
reflect only the random
uncertainties in $(\mu_{0},\gamma,a)$ and do not include systematics
such as those caused by uncertainties in ages or in distance moduli
and reddenings. The largest of these latter errors -- the $\sim 10$ per 
cent distance error for the Fornax clusters -- results in at maximum a 
$\sim 2$ per cent error in $\mu_0$ and a $\sim 20$ per cent error in 
$L_{\infty}$ -- both within the random uncertainties for these values.

\section{The distribution of core radii}
\label{discussion}
In Papers I and II, we presented evidence for
a trend in core radius with age for the cluster systems of both the 
LMC and SMC -- namely that the spread in $r_c$ increases dramatically
with increasing age. In the context of identifying the cause of
this trend, it is important to observe whether it exists uniquely in
the Magellanic Clouds (which are themselves fairly atypical members 
of the Local Group) or whether it extends to cluster systems in other 
galaxies. It is therefore very useful to take the results we have 
presented here for the Fornax and Sagittarius dSph globulars and compare 
them with those for the oldest clusters in the Magellanic systems. 

\begin{figure}
\includegraphics[width=0.5\textwidth]{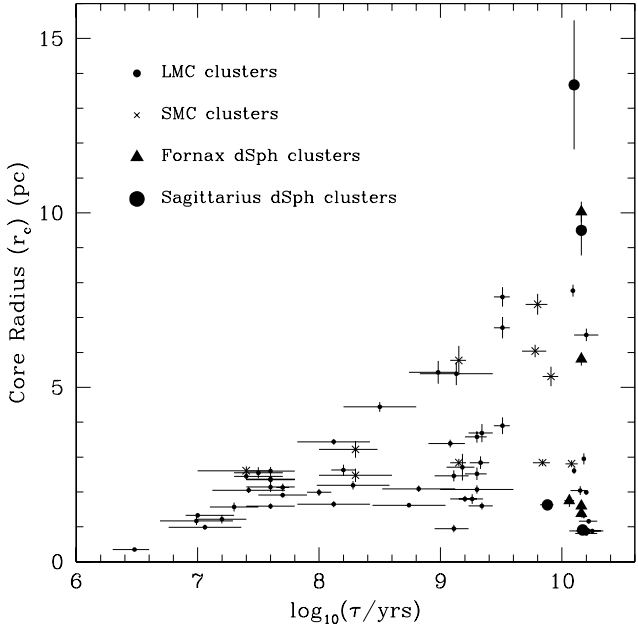}
\caption{Core radius vs. age for the 53 LMC and 10 SMC clusters from Papers I and II (small filled circles and small crosses respectively) together with the Fornax and Sagittarius dSph globular clusters from the present paper (large filled triangles and large filled circles respectively). Points with no visible error-bars have uncertainties smaller than their symbol size. Apart from Arp 2, the spread in core radii for the oldest clusters is in good agreement between each sample. Arp 2 matches the LMC cluster Reticulum (not plotted), which has $r_c \sim 14.6$ pc (see text).}
\label{radage}
\end{figure}

An immediate qualitative comparison is possible by simply adding the 
Fornax and Sagittarius clusters to the radius-age plots from Papers I 
and II (Fig. \ref{radage}). It is clear that the new clusters agree 
closely with the distribution observed for the Magellanic clusters, 
apart from Arp 2. Even this cluster is not unique however -- it is well 
matched by the LMC cluster Reticulum (see below), which was not measured 
in the sample of Paper I. 

It is worth noting briefly the result of de Grijs et al. 
\shortcite{richard} who have demonstrated that the radius-age distribution
persists even if only stars in the mass range $0.8-1.0 M_\odot$ are
used to construct the radial profiles. This means that the reduced
spread in core radius for the youngest clusters cannot be a result
of their profiles being dominated by a few high luminosity mass-segregated
stars. Similarly, de Grijs et al. show that the degree of mass
segregation for LMC clusters {\em even of different ages} is very similar.
In terms of our profile construction technique, this means that 
the omission of very faint (and in some cases very bright) stars from
the star counts should not alter the derived structural parameters
significantly (apart from a zero-point shift), and hence does not affect
any analysis in terms of the radius-age trend. We are also justified
in directly comparing the profiles of clusters of all ages.

A more detailed examination than that from the simple radius-age plot can 
be obtained by constructing cumulative distributions in $r_c$ for the
Sagittarius clusters, the Fornax clusters, and the old Magellanic Cloud
clusters. When matching these distributions against each other, it is 
important that we 
have cluster samples which are as complete as possible, otherwise 
selection effects may bias the results. We are confident that our sample 
of Fornax clusters is complete, and if we add Pal 12 to our group of 
Sagittarius clusters, this sample is also complete to the limits of
present knowledge (although it is certainly possible that other outer 
halo globular clusters will be identified as former Sagittarius members 
in the future). Unfortunately, Pal 12 is, like Terzan 7, Terzan 8, 
and Arp 2, a very sparse cluster with a distinctly patchy appearance. 
Its structure therefore appears difficult to quantify, and there is a 
wide variation in literature measurements of $r_c$. Webbink 
\shortcite{webbink} lists $r_c \sim 28\arcsec$, while Chernoff \& 
Djorgovski \shortcite{chernoff} measure $r_c \sim 31\arcsec$. In 
contrast, Trager et al. \shortcite{tragerb} have $r_c \sim 1\farcs7$,
determined from a very ragged profile. Harris \shortcite{harris} lists 
$r_c \sim 12\arcsec$ which is presumably some average of all 
available literature measurements. In the absence of other information, 
we will adopt this value -- at Harris's listed distance of 19.1 kpc it 
converts to $r_c \sim 1.1$ pc. Clearly, better measurements of this 
object are required, especially given its recently elevated status as a
Sagittarius dSph cluster.

We also need to ensure that our samples of old Magellanic clusters are 
complete. In comparison with the youngest clusters in the Fornax and
Sagittarius dSph systems (Terzan 7, $\sim 7.5$ Gyr; Pal 12, $\sim 7-8$
Gyr \cite{buonages,rosenberg,salaris}), we define ``old'' to include
all clusters with $\tau \ge 7$ Gyr. This means that in addition to
the twelve old LMC clusters from Paper I, we must definitely add the
Reticulum cluster, which is coeval with the metal poor Galactic
globulars (e.g., Walker \shortcite{walker}) and has 
$r_c = 60 \pm 20\arcsec$ \cite{webbink}, corresponding to 
$r_c = 14.6 \pm 4.5$ pc
at a distance modulus of 18.5. We must also add ESO 121-SC03, the only 
cluster to lie in the LMC age gap, with $\tau = 9 \pm 2$ Gyr and 
$r_c = 34 \pm 5\arcsec$ \cite{mateoeso}, which is $r_c = 8.3 \pm 1.2$ pc
at the LMC distance. It is also necessary to mention NGC 1928 and 
NGC 1939, which are compact clusters in the LMC bar. The spectroscopic 
study of Dutra \shortcite{dutra} has shown that these two are likely 
to be old; however no CMDs or surface brightness profiles appear in
the literature for either. For the moment therefore, 
we will not include them in our calculations -- as we will see below,
their presence or otherwise is not significant.

Unfortunately, the SMC cluster system is not as easy to quantify as
the LMC system, partly because there is no age gap for the SMC clusters. 
In addition, as discussed in Paper II, the SMC clusters have not been
extensively studied, and literature measurements of cluster ages and
structural parameters are few and far between. The sample of Paper II
included the only bona fide old SMC cluster (NGC 121), as well as 
four clusters with ages of $6-8$ Gyr (NGC 339, NGC 361, NGC 416, Kron 3).
In addition, Mighell, Sarajedini \& French \shortcite{mighell} have
shown that the cluster Lindsay 1 has $\tau \sim 9$ Gyr, and that
Lindsay 113 has $\tau \sim 5-6$ Gyr, while Piatti et al. 
\shortcite{piatti} show that Lindsay 38 has $\tau \sim 6$ Gyr. None of 
these clusters have high resolution brightness profiles in the 
literature; nor is it clear how many other massive SMC clusters might 
have $\tau \sim 7$ Gyr or greater. We are therefore unable to include 
the SMC clusters in the quantitative discussion below -- once again the 
need for extra measurements of the clusters in this system is evident.

\begin{figure}
\includegraphics[width=0.5\textwidth]{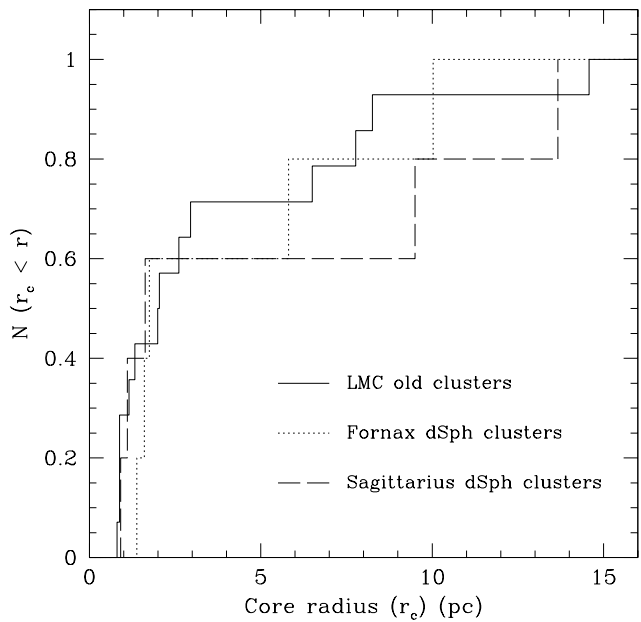}
\caption{Cumulative distributions in $r_c$ for old LMC clusters, and clusters in the Fornax and Sagittarius dwarf galaxies. Measurements for the Fornax clusters are taken from the present study, as are those for the Sagittarius clusters (except Pal 12 which is from the literature). Measurements for twelve of the fourteen LMC clusters are from Paper I; the remaining two are taken from the literature, as discussed in the text.}
\label{cumdist}
\end{figure}

The cumulative distributions in $r_c$ for the LMC old clusters and the
Fornax and Sagittarius dSph clusters are shown in Fig. \ref{cumdist}. 
Even without considering the uncertainties in the literature measurements
of $r_c$ discussed above, it is clear that the distributions match well,
especially given the small sample sizes for the dSph clusters.
There are several aspects worth considering in greater detail. Firstly,
it should be noted that for the most compact Fornax clusters, our
core radius measurements are upper limits because of the resolution
of our profiles. This means that the agreement in the distributions
at small $r_c$ is probably even better than shown. Similarly, we note
that the measurements of $r_c$ for the most compact LMC clusters from
Paper I are also upper limits. Secondly, in Paper I we discussed the
possibility that the radius-age distribution showed a bifurcation
at $\sim 100$ Myr, so that the old clusters follow an almost bimodal
distribution -- that is, they can either be compact ($r_c \le 3$ pc) 
or extended ($r_c \ge 7$ pc). We can see this distribution in the
cumulative profile for the LMC clusters, as a flat region between
$\sim 3$ and $\sim 7$ pc. It is possible that this is simply a small
sample effect. However, it seems that both of the dSph cluster systems
follow a similar distribution, with a dearth of clusters in the range
$2 \le r_c \le 6$ pc. We also note that our measurements from Paper II
show that the oldest SMC clusters also appear to have a grouped 
distribution, although again, is it likely that this sample is incomplete.

In terms of a direct comparison between the LMC and dSph cluster
distributions, we can use the apparent grouping of clusters for a
quantitative estimate. From Fig. \ref{cumdist} it is clear that 
$\sim 65-70$ per cent of old LMC clusters have $r_c < 3$ pc. This leads 
us to expect that in each dSph system, at most $0.7 \times 5$ clusters, 
or $\sim 3$ will have $r_c < 3$, which is exactly what we observe. 
A K-S test can provide a more rigorous assessment of how well the 
distributions match, although we note that such a test is not necessarily
well suited to samples as small as those we are dealing with here. 
Nonetheless, applying a K-S test to the LMC and Fornax samples
shows that they were drawn from the same distribution at $\sim 97.5$ per
cent significance. If we ignore differences between the distributions
for $r_c < 1.6$ pc (because of our resolution limits) this increases
to $> 99$ per cent. Comparing the LMC and Sagittarius samples yields
a match at $\sim 75$ per cent significance, with the largest difference
occurring between the two distributions near $r_c = 9$ pc -- a region
where stochastic effects (and measurement errors) are large. Finally,
comparing the Fornax and Sagittarius samples shows that they are similar
at $\sim 70$ per cent significance. If we again neglect the differences
for $r_c < 1.6$ pc, the significance increases to $\sim 99$ per cent.
We note that adding one or two clusters to the small $r_c$ end of
the LMC distribution (e.g., NGC 1928 and/or NGC 1939) would decrease
the significances of the match between the LMC and dSph samples 
slightly, but the agreement would remain good. Similarly, increasing
the core radius of Pal 12 to $\sim 30\arcsec$, or $\sim 3$ pc (as 
discussed above), again would not influence the agreement between the
distributions significantly. The greatest differences in Fig. 
\ref{cumdist} always occur for large $r_c$, especially given the 
resolution limits, and this region is effectively unchanged by the
addition, removal, or shifting of a couple of clusters at the low
end of the plot. 

We conclude therefore that the distribution of old clusters we observed
for the LMC in Paper I, and to a certain extent for the SMC in Paper II,
is not unique to the Magellanic systems but rather appears universal,
at least in the satellite galaxies of the Milky Way. Not only do all the 
cluster systems we have observed possess expanded clusters, they possess 
them in the same proportion, irrespective of the mass of the host galaxy,
and its isolation (or otherwise). While this does not show that the
radius-age trend of the Magellanic clusters is universal, it does 
demonstrate that the end-point of this trend matches the end-points of 
whatever paths cluster evolution has followed in the Sagittarius and 
Fornax dwarf galaxies. Given the differences between the environments 
local to these two galaxies (Fornax is relatively isolated, while 
Sagittarius is engaged in strong tidal interactions with the Milky Way) 
and that of the more massive LMC, it is clear that external influences 
are unlikely to be the driving force behind the radius-age trend, in 
agreement with the simulations of Wilkinson et al. \shortcite{mark}. If 
changing formation conditions are the key to the puzzle, as suggested by 
these authors, the implication from the present work is that these 
conditions were the same across the entire local region at the epoch of 
initial cluster formation. Given that the first burst of cluster 
formation does appear to have occurred simultaneously across both dwarf 
spheroidals, the Milky Way, and the LMC, this is not an outlandish 
suggestion. New $N$-body cluster evolution calculations are of course 
required to further explore the radius-age trend and its end-points, as 
well as the possible effects of varying formation conditions. Such 
simulations are presently in progress. Similarly, it is important to 
investigate the SMC system in more detail observationally, as described 
above. Finally, by far the largest sample of local globular clusters 
belongs to the Milky Way. An analysis of the structures of these clusters 
in the context of what we have observed for the external systems is currently 
underway.

\section{Summary and Conclusions}
We have presented surface brightness profiles for all the globular clusters 
in the Fornax and Sagittarius dwarf galaxies, derived from archival 
{\em HST} WFPC2 observations. The profiles were constructed similarly 
to those for the LMC and SMC clusters we studied in Papers I
and II, with only minor modifications to the procedure to account 
for special properties of the present data. From the surface brightness 
profiles, we have determined structural parameters for each cluster, 
including their core radii and estimates for their total luminosities 
and masses. These data, along with the surface brightness profiles are 
available on-line at 
{\em http://www.ast.cam.ac.uk/STELLARPOPS/dSph\_clusters/}. Our 
measurements of core radii are generally in reasonable agreement with
previous lower resolution measurements in the literature. However,
we have found the core radius of Terzan 7 to be smaller than previously
determined, and that for Terzan 8 to be somewhat larger. We have also
presented evidence that Fornax cluster 5 is a post core-collapse
candidate.

Examining the two cluster systems in the context of the radius-age
trend which we highlighted in Papers I and II, we find that the
distribution of cluster core radii in both dwarf galaxies matches
that for the old LMC clusters within the limits set by measurement
errors and the small sample sizes. While this is not evidence for
the dSph clusters having evolved via the same radius-age trend we
observed for Magellanic Cloud clusters, it does indicate that the
end-points of the structural evolution of clusters in these four
systems are equivalent. The fact that this equivalence exists even
though the four parent galaxies have very different masses and formation
histories, and have been influenced by the Milky Way to varying degrees,
suggests that it is not strong external influences which have primarily
determined these evolutionary end-points. By inference,
the radius-age trend in the Magellanic clusters is therefore
unlikely to be driven by external forces. 

\section*{Acknowledgments}
ADM would like to acknowledge the support of a Trinity College ERS 
grant and a British government ORS award.
This paper is based on observations made with the NASA/ESA 
{\em Hubble Space Telescope}, obtained from the data archive at the 
Space Telescope Institute. STScI is operated by the association of 
Universities for Research in Astronomy, Inc. under the NASA contract 
NAS 5-26555.

% End of paper here

%%%%%%%%%%%%%%%%%%%%%%%%%%%%%%%%%%%%

\bsp % ``This paper has been produced using the ...''

\label{lastpage}

\end{document}